\documentclass[acmlarge,nonacm]{acmart}
\usepackage{amsmath,amsfonts}
\usepackage{array}
\usepackage{subcaption}
\usepackage{textcomp}
\usepackage{url}
\usepackage{verbatim}
\usepackage{graphicx}
\usepackage{tikz}
\usepackage{float}
\usepackage{makecell}

\usepackage{xcolor}
\usepackage[ruled,vlined,linesnumbered]{algorithm2e}
\usepackage{tabularx}
\usetikzlibrary{arrows.meta}
\usetikzlibrary{positioning}
\usetikzlibrary{calc}
\usepackage{booktabs}

\AtBeginDocument{%
  }

\setcopyright{none}

\begin{document}

\title{S\emph{i}NMULI: Novel Signed Network Approach for Malicious URL Identification}

\author{Avijit Gayen}
\authornote{Also with Techno India University, West Bengal, Kolkata, India}
\email{avijit.gayen@iitg.ac.in}
\orcid{0000-0002-7187-646X}
\author{Sayan Mondal}
\email{sayan.mondal@iiitg.ac.in}
\author{Angshuman Jana}
\email{angshuman@iiitg.ac.in}
\affiliation{%
  \institution{Indian Institute of Information Technology, Guwahati}
  \city{Guwahati}
  \state{Assam}
  \country{India}
}

\renewcommand{\shortauthors}{Gayen et al.}

\begin{abstract}
In today's era of rapid advancements in artificial intelligence, computer security, and online safeguarding measures have undergone significant improvements. However, malicious websites continue to facilitate the spread of phishing schemes, fraudulent activities, and unsolicited communications. Conventional methodologies in machine learning, deep learning, and detection technologies for counterfeit websites predominantly depend on static data analysis, which frequently proves ineffective against the onslaught of malicious online entities. In response to these challenges, in this work, we propose a signed network-based approach for malicious URL identification(S{\em i}NMUlI). We introduce an innovative framework that conceptualises the identification of harmful URLs as a signed network-based binary classification problem that is strongly rooted in the fundamental principles of social network analysis and social balance theory. In this approach, a signed network is constructed based on the back-links, i.e., external hyperlinks of URLs, wherein each node symbolizes a URL and the hyperlinks function as signed edges. Utilising a balance-theoretic inference mechanism, our methodology propagates edge signs and classifies unlabeled domains by employing a 51\% majority rule across incoming links. Experimental results on this real-world dataset demonstrate that S{\em i}NMULI achieves 99.89\% accuracy, 99.62\% precision, and 99.80\% F1-score, which outperforms traditional ML and deep learning baseline models. Beyond high accuracy, S{\em i}NMULI offers interpretability, resilience against adversarial obfuscation, and independence from training data, making it a lightweight and scalable solution for real-world cyber defence.
\end{abstract}



\received{24 February 2024}
\received[revised]{12 March 2024}
\received[accepted]{5 June 2024}

\maketitle

\section{Introduction}\label{s:intro}

\begin{table*}[t]
\centering
\caption{List of Symbols and Notations Used in S{\em i}NMULI Framework}
\label{tab:symbols}
\renewcommand{\arraystretch}{1.15}
\setlength{\arrayrulewidth}{0.6pt}
\begin{tabular}{|p{2.4cm}|p{13cm}|}
\hline
\textbf{Symbols} & \textbf{Description} \\
\hline
$U_0$ & Seed set of input URLs from which crawling and hyperlink extraction begin. \\
\hline
$u$ & A specific URL under consideration during crawling or analysis.\\
\hline
$V$ & Set of nodes in the signed network graph, where each node represents a URL. \\
\hline
$V'$ & Set of structurally valid URLs retained after filtering and preprocessing. \\
\hline
$n$ & Number of initial URLs in the dataset ($n = |U_0|$). \\
\hline
$n'$ & Number of structurally valid URLs retained after cleaning ($n' = |V'|$). \\
\hline
$E$ & Set of directed edges in the signed network graph, where each edge represents a hyperlink between two URLs. \\
\hline
$e_{uv}$ & Directed edge from URL $u$ to URL $v$. \\
\hline
$s: E \rightarrow \{+1,-1\}$ & Sign function mapping each directed edge to either trust $(+1)$ or distrust $(-1)$. \\
\hline
$s_{uv}$ & Sign of the directed edge from URL $u$ to website/URL $v$; $+1$ denotes trust, $-1$ denotes distrust. \\
\hline
$L_{\text{int}}(u)$ & Set of internal hyperlinks extracted from $u$ that point to the same domain or subdomain. \\
\hline
$L_{\text{ext}}(u)$ & Set of external hyperlinks extracted from $u$ that point to different domains. \\
\hline
$E^+_v$ & Set of incoming positive (trust) edges to URL $v$. \\
\hline
$E^-_v$ & Set of incoming negative (distrust) edges to URL $v$. \\
\hline
$y_v$ & Final label assigned to URL $v$: $+1$ (legitimate) or $-1$ (malicious). \\
\hline
$T$ & Set of all triads (three-node subgraphs) in the signed network. \\
\hline
$d_{\text{avg}}$ & Average node degree in the signed network. \\
\hline
$m$ & Average number of hyperlinks per page during crawling. \\
\hline
$L$ & Size (in characters or bytes) of the HTML page being parsed. \\
\hline
$t$ & Total number of triads present in the signed network. \\
\hline
$\mathcal{J}$ & Filtered dataset \\
\hline

$N^{-}(v)$ & Incoming neighbourhood of $v$: set of nodes $u$ with a directed edge $(u,v) \in E$.\\
\hline

$n^{+}(v)$ & Number of incoming positive edges to $v$, i.e., $|\{u \in N^{-}(v) : s(u,v)=+1\}|$.\\
\hline
$n^{-}(v)$ & Number of incoming negative edges to $v$, i.e., $|\{u \in N^{-}(v) : s(u,v)=-1\}|$.\\
\hline
$e_{?}$ & Unlabeled Edge. \\
\hline

TriadsByEdge & Mapping from an edge to the list of triads that include it.\\
\hline
Majority & Threshold for labeling via incoming signs (default 0.51).\\
\hline
TiePolicy & Policy choice when the majority condition is not met (abstain, malicious, benign).\\
\hline
\end{tabular}
\end{table*}

The Internet has revolutionised how we connect, but it has also expanded the attack surface. Three features make it difficult to secure: (a) its {\em distributed, heterogeneous} design leaves no single authority to enforce uniform defences across domains \cite{schneier2015data}; (b) a {\em non-unified security regime} across Autonomous Systems means gaps at their boundaries can be exploited \cite{anderson2010security}; and (c) {\em anonymous} access lowers accountability for malicious actors \cite{saleem2022anonymity}. Together, these conditions enable a wide range of attacks---malware, ransomware, phishing, adware, and malvertising---aimed at data theft, service disruption, and fraud. In practice, malicious URLs are a common entry point: they deliver payloads, redirect to fraudulent sites, or coax users into disclosing credentials.

\par
\begin{figure*}[t]
    \centering

    \begin{subfigure}{0.48\textwidth}
        \centering
        \includegraphics[width=\linewidth, keepaspectratio]{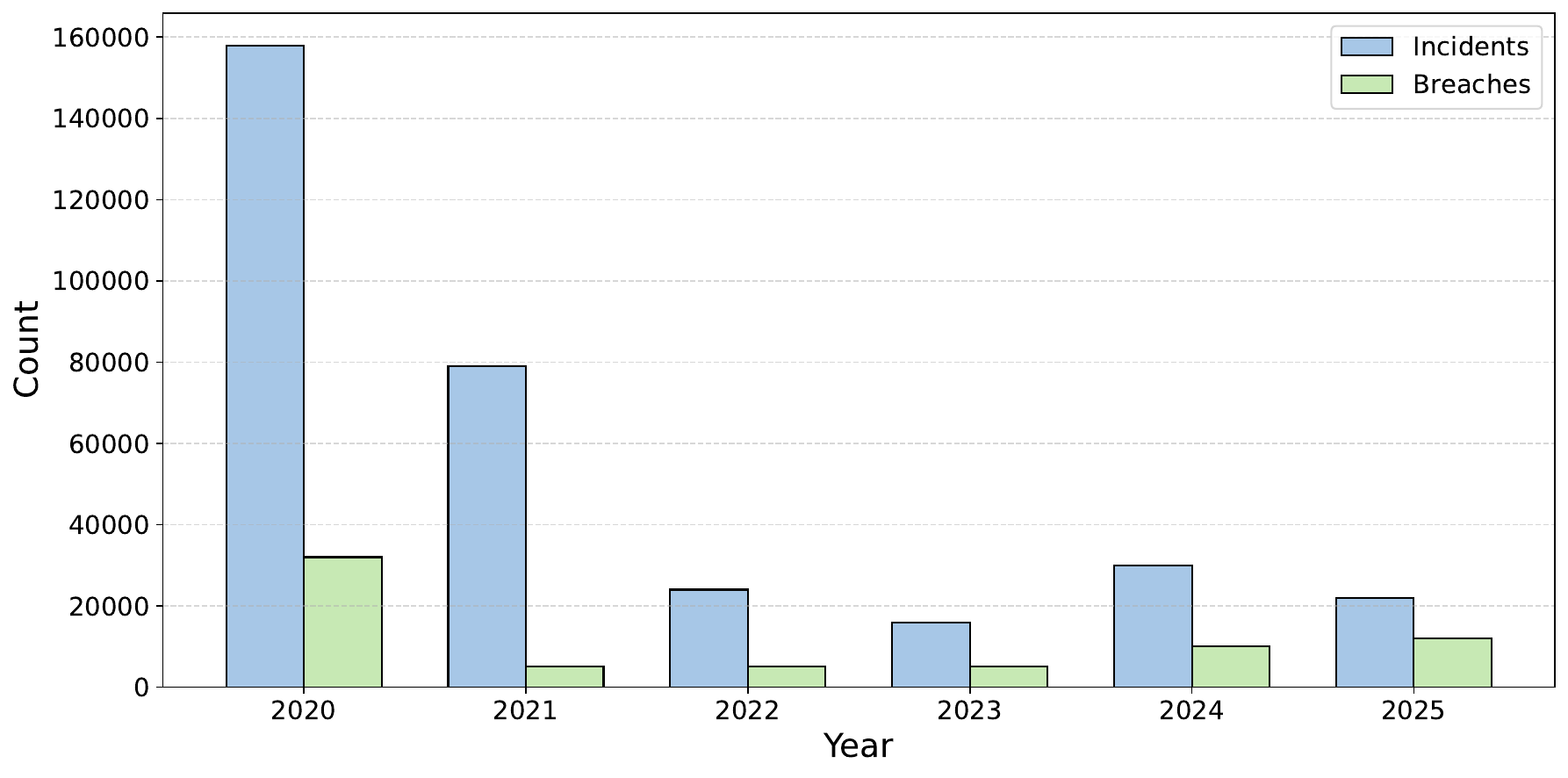}
        \caption{Statistics of cyberattacks (incidents vs.\ confirmed breaches) between 2020--2025.}
        \label{fig:growth}
    \end{subfigure}
    \hfill
    \begin{subfigure}{0.48\textwidth}
        \centering
        \includegraphics[width=\linewidth, keepaspectratio]{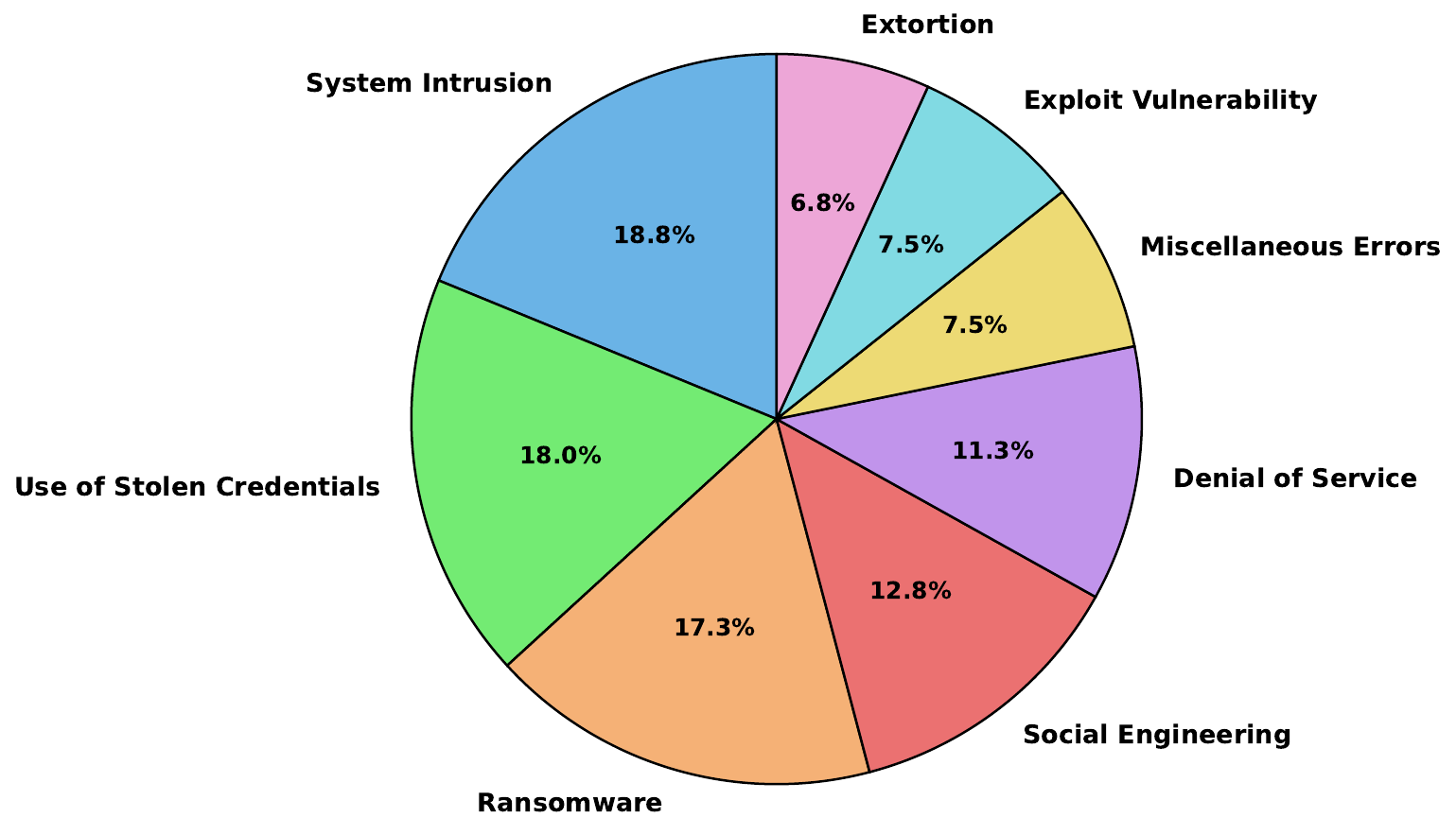}
        \caption{Figure shows the top 8 cyberattack types in 2025 in a pie chart; major vectors are shown in proportional distribution.}
        \label{fig:attacktypes}
    \end{subfigure}
\label{fig:cyberattack_stat}
\caption{Figure shows the cyber attack statistics over a period of time(2020-2025) using a bar graph, and types of cyber attacks with their share are represented in a pie chart.}
\end{figure*}

\begin{figure}[h!]
    \centering
    \includegraphics[width=\linewidth]{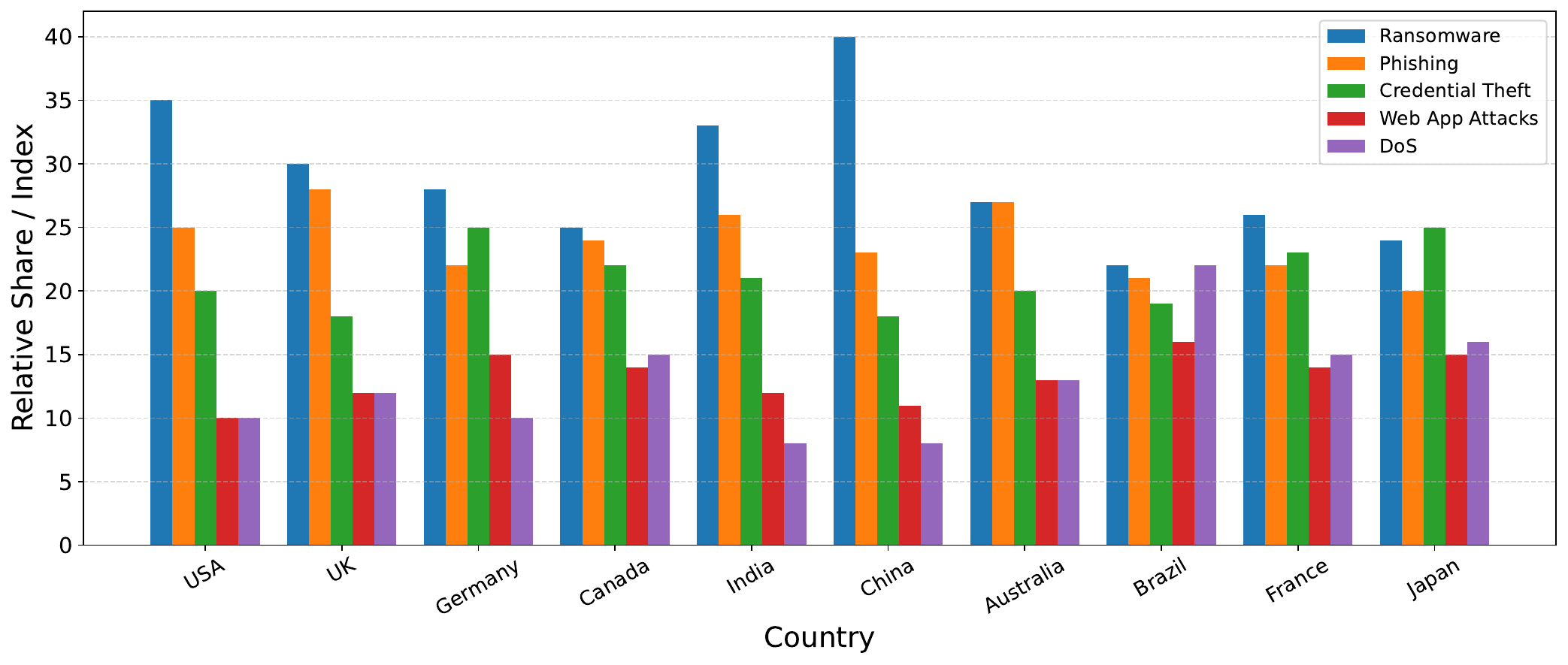}
    \caption{Figure shows the distribution of the top five attack types across the ten most affected countries in 2025. Along the s-axis, we represent the countries, and along the y-axis, we represent the no. of incidents that occurred.}
    \label{fig:countries}
\end{figure}

 The scale of the problem is substantial. In Q2 2023 alone, reported phishing and other malicious incidents exceeded 1.2 million, with finance the most targeted sector (23.5\%) \cite{apwg2023}. In India, cybercrime cases rose by 11.8\% in 2020, reaching 50{,}035 incidents, largely driven by fraud, extortion, and exploitation \cite{ncrb}. Data from 2020--2025 shows steady growth in both the number and the variety of attacks \cite{verizon2025dbir}. As Fig.~\ref{fig:growth} shows, confirmed breaches hit a record {\bf 12{,}195 in 2025}, even though the overall incident count fluctuated. The mix of attack types has also shifted. Fig.~\ref{fig:attacktypes} highlights \textit{System Intrusion (18.8\%)}, \textit{Use of Stolen Credentials (18.0\%)}, and \textit{Ransomware (17.3\%)} as the leading categories, with \textit{Social Engineering (12.8\%)} and \textit{Denial of Service (11.3\%)} still playing a major role. The continued prominence of credential theft and other human-focused attacks shows that users remain exposed even when technical defenses are in place. We also observe that different types of cyber attacks are concentrated in different geographical locations (as shown in Fig.~\ref{fig:countries}). Ransomware is more common in North America and Western Europe, while credential abuse and denial-of-service attacks appear more frequently in Asia--Pacific and Latin America. Overall, the threat landscape is both growing and becoming more diverse, which calls for a mix of strong technical controls, user-focused measures, and region-specific policy responses.

There exists a plethora of works related to malicious URL detection using existing White-blacklist method ~\cite{Khonji2013Survey,su151813369}, lexical or content-driven heuristics techniques~\cite{joshi2019}, machine learning~\cite{sahoo2017, patgiri2021}, deep learning~\cite{le2018} methods. The hyperlink relationship among the URLs not only reveals the connectivities among the URLs that drive the network traffic flow among the URLs, but also reflects the intentions of creation of those hyperlinks.  We find very few works~\cite{zhou2023, gao2020,Gayen2024} that address this issue of malicious URL detection using the hyperlink relationship among the URLs from the perspective of a complex network. Thus, this strong need drives us to develop such a framework, which can address the issue from a complex network perspective.

 \par In contrast to traditional ML or heuristic methods, our approach leverages a signed hyperlink-based network structure, where URLs are modelled as nodes and hyperlinks as directed signed edges. This network-centric view provides a relational perspective and enables structural analysis using social theories to classify URLs effectively. Traditionally, from the social theories perspective, a signed network is a graph where edges between the social entities represent the nature of the relationships marked by either positive (+) or negative (-). In our context, we hypothesize that hyperlink relationships between URLs always have either a positive or a negative aspect, which reveals the intuition behind the back-link generation by the URL. We consider that where back-links originate from malicious URLs are encoded as (-) edges and the back-links emerge from the legitimate URLs represented as (+) \cite{gibbs2018introduction}. Throughout this work, ``URL'', ``domain'' and ``website'' are used interchangeably, as each node in the constructed signed network corresponds to a unique domain or subdomain. Multiple URLs belonging to the same host are aggregated under a single website node for structural analysis.
\par In our earlier work~\cite{Gayen2024}, we have proposed a network representation of this hyperlink relationship among the URLs, which emerges as a complex network. In this network, we have identified three categories of node: a) Legitimate b) Malicious c) Unlabeled URLs. In this work, we propose a Malicious URL identification algorithm based on the well-known Balance theory~\cite{cartwright1956structural}. The main objective of the proposed algorithm is to reveal the sign of the hyperlinks of many unlabeled URLs. Based on the labeling of the sign of the hyperlink, we further classify unlabeled URLs into malicious or legitimate. Building upon earlier works that focus on content-based or lexical URL features, we develop a signed network-based approach that exploits the underlying hyperlink structure among websites. This approach enriches contextual understanding and enables robust classification even for previously unseen or unlabeled domains. However, the underlying structure (semantics of the websites) can not be manipulated to a large extent by the attackers.

\par  To lead our investigation, we develop a series of research questions that describe the scope and objectives of this study.

\begin{itemize}
    \item {\bf RQ1:} {\em Can malicious URL identification be effectively modeled as a signed network classification problem, over the existing methods?}
    \item {\bf RQ2:} {\em How much the signed hyperlink network model is computationally improved model compared to existing models?}
    \item {\bf RQ3:} {\em To what extent can proposed directed balance theory improve inference of unknown URLs and enhance classification accuracy in signed hyperlink networks?}
\end{itemize}

\par These research issues serve as the cornerstone of our approach and drive the application of network-theoretic reasoning to malicious URL identification. In cybersecurity, the terms \textit{detection} and \textit{identification} are often mentioned in conjunction, yet they denote fundamentally different objectives. \textbf{Detection} refers to the act of recognising whether a given entity (e.g., a URL) exhibits malicious behaviour. It is a decision-making process, typically binary, that determines the presence or absence of a threat. \textbf{Identification}, by contrast, extends beyond the recognition of a threat to reveal its specific nature, origin, or role in a larger attack strategy. Identification often seeks to attribute the source of an attack, profile adversarial behaviour, or trace the campaign structure behind malicious activities. In our work, we specifically perform malicious URL identification. We analyse the behaviour of the malicious URLs in the proposed network to identify their distinct behaviour in the hyperlink network that has been discussed in detail in section~\ref{s:mal_analysis}.

\par To construct the proposed network model, we crawled a benchmark dataset and extracted hyperlink structures, internal and external links, and page-level content. Further applying balance theory on the proposed signed hyperlink network, we inferred the sign of unlabeled edges of those back-links of the unknown URL and classified previously unlabeled URLs. Preliminary results reveal that malicious websites show lower centrality scores, fewer internal links, and sparse clustering coefficients compared to legitimate ones, supporting the feasibility of our network-based detection approach. Specifically, it achieves a classification accuracy of 99.89\%, surpassing the best-performing baseline by a margin of 2.4\%, while also recording a precision of 99.62\% and F1-score of 99.80\%, indicating superior reliability in both minimizing false positives and capturing true malicious instances.

\par Although hyperlink structures have the ability to differentiate between malicious and legitimate websites, there were a number of significant obstacles in the data collection process.  First, trustworthy crawling and data extraction are hampered by the short operational lifespan of many malicious URLs and their use of evasive techniques like bot detection, CAPTCHA, or restricted access (e.g., HTTP 403/404 errors).  As a result, many URLs were either unparsable or produced hyperlink structures that were insufficient.  Second, it is challenging to create hyperlink structure because malicious websites' dynamic and unstable nature frequently results in sparse or out-of-date linking behavior.  Third, the automated extraction of internal and external web pages was made more difficult by user interaction dependencies on contemporary websites, such as login-gated content or links rendered by JavaScript. These restrictions limited the overall pool of usable domains and required strict filtering.  However, it reaffirmed how crucial it is to use structural patterns rather than conventional content-based signals when identifying malicious URLs.  The lack of hyperlink information in the benchmark data set necessitated custom web scraping and filtering in order to create a signed, meaningful network, which was a major challenge.
Despite these limitations, our study contributes a new angle to malicious detection by:

\begin{itemize}
    \item We propose a novel approach to identifying malicious URLs that models websites and hyperlinks as a signed, directed network, leveraging social balance theory for structural inference.
    
    \item We propose a directed balance model that can reliably classify unlabeled nodes without requiring large amounts of training data.
    
    \item Experiments on big benchmark datasets show that the proposed model has almost perfect detection ability, and it also provides clear explanations, is strong against tricky hidden attacks, and can handle large amounts of data.
\end{itemize}

\par This paper is organised as follows. Section~\ref{s:related} reviews prior work on malicious URL detection and discusses key limitations in existing approaches. Section~\ref{s:background} presents the theoretical background of our approach, with a focus on social balance theory in signed networks. Section~\ref{s:method} describes the proposed methodology, including data pre-processing, network construction, and the classification algorithm. Section~\ref{s:result} reports and analyses the experimental results, while Section~\ref{s:mal_analysis} examines the behaviour and characteristics of malicious URLs in more detail. Finally, Section~\ref{s:conclusion} concludes the paper and outlines directions for future research. To enhance the readability of the paper, we have defined all the notation and symbols in the table~\ref{tab:symbols} at the very beginning of the paper.

\section{Related Works}
\label{s:related}

Research on malicious URL detection has gradually shifted from simple list-based checks to more complex, network-aware approaches. Below, we briefly outline this evolution in roughly chronological order and highlight representative work along the way.

Early systems relied on \textit{blacklists} and \textit{whitelists}, where each URL was checked against collections of known malicious or trusted entries. Services such as Google Safe Browsing and PhishTank helped popularise this strategy by maintaining continuously updated feeds~\cite{Khonji2013Survey, su151813369}. These methods work well for previously reported threats and offer fast lookups, but they perform poorly at \textit{zero-hour}~\cite{o2013zero}. Studies report low coverage at the time of attack, since many phishing campaigns evolve quickly or disappear within hours, often before list updates can propagate~\cite{kumar2020novel, bell2020analysis}.

To move beyond pure lookup, researchers examined what a URL \emph{looks} like: length, token patterns, special characters, and suspicious keywords. McGrath and Gupta~\cite{mcgrath2008} documented characteristic differences between phishing and benign URLs, and Joshi et al.~\cite{joshi2019} showed that ensembles over static lexical features can be accurate without crawling or list queries. The catch is brittleness: simple obfuscation (URL shortening, token insertion, homoglyphs) can defeat fixed rules.

Supervised and unsupervised learning broadened the feature space to include lexical, host-based, and sometimes content or DNS signals. Sahoo et al.~\cite{sahoo2017} surveyed these techniques, and Ma et al.~\cite{ma2009} demonstrated that learned models improve over blacklist baselines when given richer representations (see also~\cite{RASHID2024110398}). However, classic ML pipelines still rely on feature engineering and can drift as attacker tactics evolve.

Deep models reduce manual feature design by learning representations directly from raw strings or tokens. URLNet~\cite{le2018} learns embeddings for end-to-end classification, and hybrid CNN--GRU architectures further capture local patterns and sequence dependencies~\cite{gao2020}. Patgiri et al.'s ``DeepBF''~\cite{patgiri2021} combines learned Bloom filters with evolutionary training to speed filtering. Despite strong results, these models are data- and compute-hungry, and real-time deployment must contend with latency, label scarcity, and concept drift.

Because attacks span domains, IPs, and infrastructure, graph formulations model relationships among web entities. Recent work applies GNNs to heterogeneous graphs to capture higher-order structure~\cite{zhou2023}; other work combines URL-level features with network-based signals and applies probabilistic inference methods, such as Loopy Belief Propagation, to achieve scalable detection~\cite{guo2025}. Graph-based approaches provide a more holistic view of the ecosystem but introduce their own challenges, including large-scale graph construction, handling streaming updates, and coping with noisy or unreliable edges.

Signed relations (trust/distrust) bring social-theoretic priors into play. Heider's balance theory and status theory, later operationalised by Leskovec et al.~\cite{heider1946attitudes, leskovec2010signed}, explain regularities in signed triads and hierarchical structure; they underpin tasks like sign prediction~\cite{chiang2014prediction}, trust-aware recommendation~\cite{yang2012trust}, and link recovery in noisy graphs~\cite{kou2020}, with validation on platforms such as Epinions and Slashdot~\cite{leskovec2010signed}. For security, Kim et al.~\cite{kim2022} analyse behavioural patterns over networks to resist common evasions, and Kou et al.~\cite{kou2020} model missing links under privacy constraints. This body of work motivates trust-aware, signed-graph formulations for malicious URL detection, where balance/status cues help propagate credible signals through the web of entities.

\par While recent work has applied graph learning to domain reputation and bot detection, applications of signed social theories to adversarial environments such as phishing remain sparse~\cite{tajbakhsh2020fake}. Existing approaches to malicious URL detection largely rely on blacklists, content-based ML models, or deep neural networks~\cite{patgiri2021, le2018, kim2022}. These methods, while powerful, often require labeled datasets and suffer from poor interpretability and robustness to obfuscation. Despite interpretability and success in modeling social relationships, signed networks remain underutilized in cybersecurity contexts. Thus, there is a strong requirement to explore the role of signed relationships among the URLs in malicious URLs. In this work, we propose a framework, \textbf{S{\em i}NMULI} (Signed Network-based Malicious URL Identification), that introduces a lightweight and interpretable alternative for malicious URL detection.

\section{Background}\label{s:background}

This section provides the theoretical foundations required for our proposed framework. We first motivate the need for a network-theoretic perspective in malicious URL detection, then introduce two cornerstone concepts from signed network analysis: \textit{Balance Theory} and \textit{Status Theory}.

These theories, originally developed for social networks, provide useful tools for modelling relational structure and inferring patterns of trust and distrust. While many existing URL detection methods focus on isolated features of individual URLs, we argue that incorporating these relational perspectives yields more robust and interpretable detection.

This motivation arises from clear limitations in current approaches. Lexical, content-based, and standard machine learning methods typically treat each URL independently and overlook the underlying web of relations. In practice, websites form a complex system of interlinked entities. Hyperlinks encode implicit judgments: some links signal endorsement or trust, whereas others are adversarial or manipulative, crafted to mislead users or to mask malicious intent. Capturing these nuanced relationships requires theories that can systematically model and infer the structural consistency of trust--distrust dynamics. Balance Theory and Status Theory provide precisely such mechanisms, making them highly relevant for malicious URL identification. Importantly, this section only outlines their classical formulations; our proposed modified version of the balance theory in the context of malicious URL identification from the directed hyperlink network will be discussed in Section~\ref{s:ModifiedBalanceTheory}.

\par In the next, we discuss in detail the two major social theories, i.e., a) Social Balance theory and b) Social Status theory. We outline the social theories as well as point out the rationale behind the adaptation of the social theory in the context of malicious URL identification.

\subsection{Balance Theory}
\begin{figure*}[t]
\centering

\begin{subfigure}{0.23\linewidth}
    \centering
    \includegraphics[width=\linewidth]{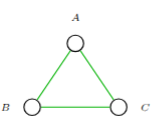}
    \caption{Balanced: $(+,+,+)$ Strong Balanced}
    \label{fig:triad-bpp}
\end{subfigure}
\hfill
\begin{subfigure}{0.23\linewidth}
    \centering
    \includegraphics[width=\linewidth]{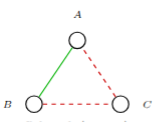}
    \caption{Balanced: $(+,-,-)$ Strong Balanced}
    \label{fig:triad-bmm}
\end{subfigure}
\hfill
\begin{subfigure}{0.23\linewidth}
    \centering
    \includegraphics[width=\linewidth]{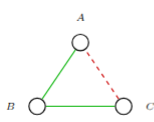}
    \caption{Unbalanced: $(+,+,-)$ Strong Unbalanced}
    \label{fig:triad-ub}
\end{subfigure}
\hfill
\begin{subfigure}{0.23\linewidth}
    \centering
    \includegraphics[width=\linewidth]{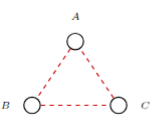}
    \caption{All negative: $(-,-,-)$ Weak Balanced}
    \label{fig:triad-allneg}
\end{subfigure}

\caption{Figure represents the possible triads in Balance-theory. The solid green undirected edges denote trust $(+)$, whereas red dashed undirected edges denote distrust $(-)$. Two triads are balanced under both strong and weak definitions \subref{fig:triad-bpp}, \subref{fig:triad-bmm}. Triad \subref{fig:triad-ub} is unbalanced under both. Triad \subref{fig:triad-allneg} is unbalanced under strong balance (Cartwright--Harary) but balanced under weak balance (Davis).}
\label{fig:balance-triads}
\end{figure*}

The balance theory, introduced by Heider (1946) and extended by Cartwright and Harary (1956), asserts that social networks evolve toward a balanced state, where triads of relationships conform to certain stability rules. In a signed network, a \emph{triad} is defined as a set of three nodes connected by signed edges representing positive ($+1$, trust) or negative ($-1$, distrust) relationships. A triad $(i,j,k)$ is considered \emph{balanced} if the product of its edge signs satisfies:

\begin{equation}
s_{ij} \cdot s_{jk} \cdot s_{ki} > 0
\end{equation}
where $s_{xy} \in \{+1, -1\}$ is the sign of the directed edge from node $x$ to node $y$. From a structural perspective, two configurations of triad yield balance:
\begin{enumerate}
    \item \textbf{Mutual trust:} All three edges are positive $(+,+,+)$.
    \item \textbf{Common adversary:} Exactly one edge is positive and the other two are negative $(+,-,-)$, indicating that two mutually distrustful nodes share a common enemy.
\end{enumerate}
In the balancing of the triad, there exist two main interpretations of balance in the literature:

\begin{itemize}
    \item \textbf{Strong Balance (Cartwright--Harary)} A triad is balanced only if it contains an \emph{even} number of negative edges (0 or 2) (as shown in fig.~\ref{fig:triad-bpp}, ~\ref{fig:triad-bmm} and ~\ref{fig:triad-ub}). Under this definition, a triad with all negative edges $(-,-,-)$ is considered \emph{unbalanced}, as it represents universal mutual distrust, which is structurally unstable.  
    \item \textbf{Weak Balance (Davis)} A triad is balanced if it contains \emph{no more than one} positive edge. This broader definition permits $(-,-,-)$ as balanced, modeling scenarios where mutual distrust can be stable, such as isolated hostile cliques (as shown in fig.~\ref{fig:triad-allneg}).
\end{itemize}
In the area of detecting harmful actions, balance theory offers a clear way to figure out missing signs on connections by making sure the relationships between three connected items are consistent. If a group of three items has one or more connections with unknown signs, the missing signs can be guessed in a way that keeps the whole setup balanced. This method is especially helpful in networks where links have some known labels and others are not, since it helps determine whether a website is bad or safe when that information isn't already known.



\subsection{Status Theory}

Status theory, introduced by Leskovec, Huttenlocher, and Kleinberg (2010  \cite{leskovec2010signed}), offers an alternative view that is well suited to directed signed networks. It assumes that trust (or esteem) is organised hierarchically: a positive edge from node \(A\) to node \(B\) means that \(A\) assigns a higher status to \(B\), while a negative edge indicates that \(A\) views \(B\) as lower in status. This framework stands strongly on the following principles---
\begin{itemize}
    \item A positive directed edge from node $A$ to node $B$ indicates that $A$ views $B$ as having \textbf{higher status}.
    \item A negative directed edge from $A$ to $B$ indicates that $A$ perceives $B$ as having \textbf{lower status}.
\end{itemize}

\begin{figure}[t]
\centering
\begin{tikzpicture}[
  scale=0.95,
  every node/.style={font=\scriptsize},
  dot/.style={circle, draw, fill=white, minimum size=6mm, inner sep=0pt, thick},
  edge/.style={line width=0.9pt, shorten >=1.3mm, shorten <=1.3mm},
  posedge/.style={edge, -{Stealth[length=2.2mm,width=1.6mm]}, draw=green!70!black},
  negedge/.style={edge, -{Stealth[length=2.2mm,width=1.6mm]}, draw=red!80!black, dashed},
  title/.style={
    font=\footnotesize\bfseries,
    fill=white, rounded corners=1pt, inner sep=1.2pt,
    text width=1.4cm, align=center
  }
]

\newcommand{\IsoTriad}[3]{
  \begin{scope}[xshift=#1]
    \coordinate (A) at (0,1.05);
    \coordinate (C) at (1.35,-0.75);
    \coordinate (M) at ($ (A)!0.5!(C) $);
    \coordinate (B) at ($ (M)!0.85!-90:(C) $);
    \node[dot] (A_node) at (A) {A};
    \node[dot] (B_node) at (B) {B};
    \node[dot] (C_node) at (C) {C};
    #3
    \node[title, below=7pt of C_node] {#2};
  \end{scope}
}

\IsoTriad{-3.6cm}{Case 1}{
  \draw[posedge] (B)--(A);
  \draw[posedge] (C)--(A);
  \draw[negedge] (B)--(C);
}

\IsoTriad{0cm}{Case 2}{
  \draw[posedge] (C)--(B);
  \draw[posedge] (C)--(A);
  \draw[negedge] (A)--(B);
}

\IsoTriad{3.6cm}{Case 3}{
  \draw[negedge] (A)--(B);
  \draw[negedge] (A)--(C);
  \draw[negedge] (B)--(C);
}

\end{tikzpicture}
\caption{The figure represents a few cases of directed signed triads as per the status theory. The Green solid arrows denote positive $(+)$ ties, while red dashed arrows denote negative $(-)$ ties. 
}

\label{fig:status-theory}
\end{figure}
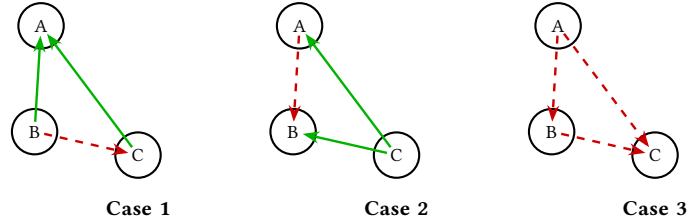

This approach captures asymmetrical relationships, making it well-suited for analysing online trust systems, reputation networks, or authority-based ranking structures. In hierarchical trust propagation, the aim is to position nodes along a latent ``status axis'' so that the direction and sign of edges align with these inferred rankings, as shown in figure \ref{fig:status-theory}. In the context of malicious URL identification, if a legitimate website frequently links positively to another site, it can be interpreted as elevating that site's credibility. Conversely, a negative link (e.g., a blacklist reference) indicates a lower perceived status. Thus, the intuition behind the back-link creation could be modelled well in the persuasion of status theory, which might be helpful in malicious URL identification.

\section{Proposed Balance Theory to Identify Malicious URL} \label{s:ModifiedBalanceTheory}
In the context of malicious URL detection, we use balance theory as a structural inference tool because it captures the inherent tendency of networks to evolve towards stable configurations of trust and distrust. Balance theory claims that social networks tend towards \emph{balanced states}, where the relationships in a triangle obey certain stability rules. This property is particularly relevant for malicious behavior detection, as signed network structures can be used to infer missing edge labels and detect unusual patterns.

However, classical balance theory\cite{INSKO198489} was formulated for undirected social graphs, assuming that all relationships are symmetric. 
In malicious URL networks, this assumption breaks down. Two key observations motivate our modified model of malicious URL detection, which adapts classical equilibrium ideas within a status-theoretic view:

\textbf{(A)} Malicious URLs frequently create deliberate, asymmetric outbound links to benign domains, for example, to blend in, borrow legitimacy, or load external resources. Benign domains, in contrast, rarely link back to such malicious sites, so the resulting pattern is strongly one-sided.

\textbf{(B)} The connections in our network are not all equivalent: edge direction encodes intent and influence, which is crucial in adversarial settings. Status theory offers a related viewpoint by imposing a hierarchical ordering on nodes in signed networks. However, classical status theory is defined for undirected graphs and does not naturally handle asymmetric, intent-driven edges. As a result, it cannot be directly applied to our malicious URL detection task, which is based on directed hyperlink relationships.

To address these limitations, we introduce a Directed Balance Theory that combines ideas from both equilibrium and state theory, specifically designed for directed signed graphs. Our approach keeps the structural stability rules from equilibrium theory and includes a direction-based understanding of trust and distrust, as suggested by state theory. This hybrid approach allows us to model asymmetric, malicious-to-benign and benign-to-malicious relationships accurately, enabling inference in partially labeled graphs and supporting more robust malicious URL classification.

\subsection{Proposed Directed Balance Theory}
Let $G = (V, E, S)$ be a directed signed graph, where $V$ is the set of nodes (URLs or websites), $E \subseteq V \times V$ is the set of directed edges, and $S: E \rightarrow \{+1, -1\}$ assigns each directed edge a sign representing trust $(+1)$ or distrust $(-1)$. A \emph{directed triad} $(u, v, w)$ consists of three distinct nodes connected by directed edges $e_{uv}$, $e_{vw}$, and $e_{wu}$. 
In a directed signed hyperlink network, we propose two basic hypotheses to integrate the directional asymmetry as well as the sign of the edge based on the intuition behind the development of that hyperlink---
\begin{itemize}
    \item $S_{xy} \neq S_{yx}$
    \item $S_{xy}$ is $(+1)$, if $x$ is a benign URL, otherwise $(-1)$ , if $x$ is malicious. 
\end{itemize}

\begin{figure}[t]
\centering
\begin{tikzpicture}[
  scale=0.9,
  every node/.style={font=\scriptsize},
  dot/.style={circle, draw, fill=white, minimum size=5mm, inner sep=0pt, line width=0.6pt},
  posedge/.style={-{Stealth[length=2.5mm,width=2mm]}, line width=0.7pt, draw=green!60!black},
  negedge/.style={-{Stealth[length=2.5mm,width=2mm]}, line width=0.7pt, draw=red!75!black, dashed},
  posedge_noarrow/.style={line width=0.7pt, draw=green!60!black},
  title/.style={font=\footnotesize\bfseries}
]

\begin{scope}[xshift=-2.2cm]
  \node[title] at (0,1.3) {Classical Undirected Balance};
  \coordinate (A1) at (0,0.55);
  \coordinate (B1) at (-0.85,-0.45);
  \coordinate (C1) at (0.85,-0.45);

  \node[dot] (NA1) at (A1) {A};
  \node[dot] (NB1) at (B1) {B};
  \node[dot] (NC1) at (C1) {C};

  \draw[posedge_noarrow] (NA1)--(NB1);
  \draw[posedge_noarrow] (NB1)--(NC1);
  \draw[posedge_noarrow] (NC1)--(NA1);

  \node at (0,-0.9) {\scriptsize Balanced: $(+,+,+)$};
\end{scope}

\begin{scope}[xshift=2.2cm]
  \node[title] at (0,1.3) {Modified Directed Balance};
  \coordinate (A2) at (0,0.55);
  \coordinate (B2) at (-0.85,-0.45);
  \coordinate (C2) at (0.85,-0.45);

  \node[dot] (NA2) at (A2) {Mal.};
  \node[dot] (NB2) at (B2) {Ben.};
  \node[dot] (NC2) at (C2) {Ben.};

  \draw[negedge] (NA2)--(NB2); 
  \draw[posedge] (NB2)--(NC2); 
  \draw[negedge] (NA2)--(NC2); 

  \node at (0,-0.9) {\scriptsize Balanced: $(-,+,-)$};
\end{scope}

\end{tikzpicture}
\caption{Figure represents the comparison of the classical undirected balance triad(left) and the proposed modified directed balance triad(right) for malicious URL detection. In the modified directed balance triad, green solid arrows denote trust $(+)$ and red dashed arrows denote distrust $(-)$.}
\label{fig:modified-balance-narrow}
\end{figure}
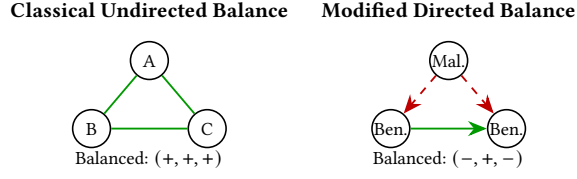

\begin{figure}
    \includegraphics[width=6.5in,height=5.5in]{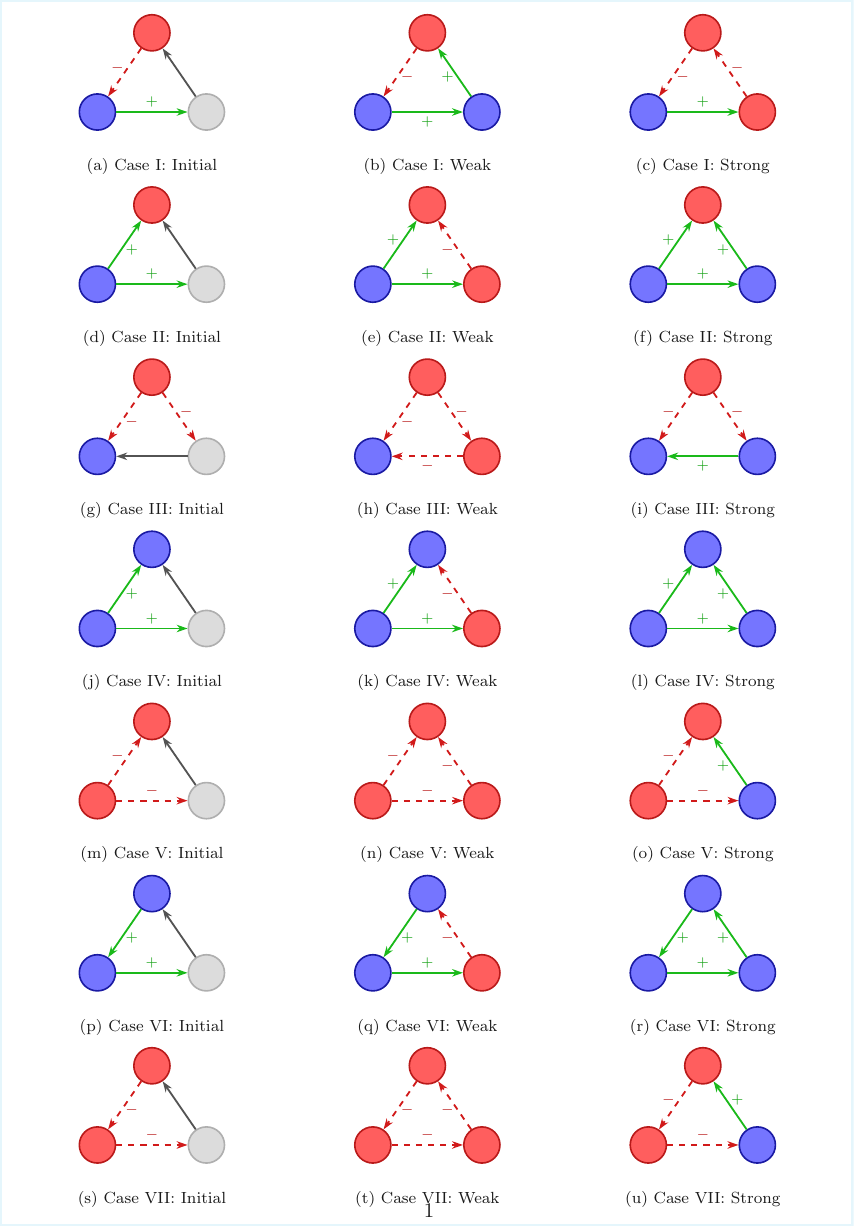}
    \caption{Figure shows the triad taxonomy across seven possible cases (I--VII). It shows the possible triad evolution under the proposed directed balance theory for malicious URL detection. Blue nodes in the triad represent the benign URL, whereas red nodes are considered to be malicious. We colored the unlabeled URL as grey, and the edges originating from those corresponding nodes have similar colors. Each row illustrates one of the seven canonical triad cases. The Initial column shows a partially signed triad. The Weak balance is shown in the second column, and the third column shows the possible strong balance.}
    \label{fig:direct_bal-cases}
\end{figure}

In figure~\ref{fig:modified-balance-narrow}, we compare the existing balance theory and our proposed directed balance theory in the context of a signed hyperlink network. We further integrate the strong and weak balance theory in our proposed directed balance theory to be used for identifying malicious URLs. In our work, we consider that we have a large set of unknown URLs to classify them into either the malicious or the benign category. Thus, we consider that there exist three categories of nodes in the network, i.e., malicious, benign, and unlabeled. The edges originate from unlabeled URLs; we consider their signs to be unknown. Based on the proposed directed balance theory, we could infer them. We apply both strong balance possibilities as well as weak balance for inference. We demonstrate all possible triad balances in our proposed model in Figure~\ref {fig:direct_bal-cases}. In the triadic relationship exploration in the proposed model, we have restricted at most one unlabeled node with exactly one unlabeled edge in a triad. This formation would help us further infer the sign of the unlabeled edge using the proposed directed balance theory.

The proposed directed balance theory integrates Structural stability from balance theory. We consider the strong as well as weak balance theory in our proposed model. We also incorporate the directional asymmetry from status theory. By merging these concepts, the proposed Modified Directed Balance Theory extends balance theory to directed systems, enabling reliable inference of missing signs and robust classification of malicious URLs in adversarial, asymmetrical network environments.

Let $G = (V, E, s)$ be a directed signed graph, where $V$ is the set of nodes (websites), $E \subseteq V \times V$ is the set of directed edges (hyperlinks),
and $s: E \rightarrow \{-1, 0, +1\}$ assigns a sign to each edge. A positive
sign $s(i,j) = +1$ denotes a trustful or benign relationship from $i$ to $j$,
while $s(i,j) = -1$ indicates a distrustful or malicious tendency. Unknown or
inferred edges are denoted by $s(i,j) = 0$.

Each node $v \in V$ is associated with a class label
$y_v \in \{\text{benign}, \text{malicious}, \text{unknown}\}$.
We use the incoming neighbourhood of $v$,
$\mathcal{N}^{-}(v) = \{u \in V : (u,v) \in E\}$,
to infer its label from edge signs.

\paragraph{Majority (51\%) rule.}
Let
\begin{align}
n_{+}(v) &= \left|\{u \in \mathcal{N}^{-}(v) : s(u,v) = +1\}\right|, \\
n_{-}(v) &= \left|\{u \in \mathcal{N}^{-}(v) : s(u,v) = -1\}\right|.
\end{align}
We assign a label to $v$ as
\[
y_v =
\begin{cases}
\text{benign}    & \text{if } \frac{n_{+}(v)}{n_{+}(v) + n_{-}(v)} > 0.51, \\
\text{malicious} & \text{if } \frac{n_{-}(v)}{n_{+}(v) + n_{-}(v)} > 0.51, \\
\text{unknown}   & \text{otherwise.}
\end{cases}
\]

\paragraph{Balance-based edge inference.}
For a triad $(i,j,k)$ with directed edges $(i,j), (j,k), (i,k)$, we denote the
known signs by $s(i,j)$ and $s(j,k)$ and infer the unknown sign $s(i,k)$ by
imposing a strong-balance constraint
\[
s(i,j) \cdot s(j,k) \cdot s(i,k) = +1,
\]
Whenever two edge signs are known. If this constraint conflicts with
existing node labels, we fall back to a weak-balance variant where only the product of signs on simple cycles is constrained to be
non-negative.

\section{Proposed Method}\label{s:method} 
In this section, we describe the proposed method of the malicious URL identification framework. Initially, we describe the benchmark dataset used as a seed dataset for further crawling to figure out the hyperlink relationship among them in our work. We analyze the dataset before crawling the URLs from their backlinks to find new URLs that are unlabeled. Further, we propose a data collection technique based on the benchmark seed dataset to build the hyperlink connection among the URLs. We also provide an analysis of the collected dataset to gain insight into the data before forming the signed network. Afterward, we formed the signed hyperlink network based on the collected dataset. We finally propose our signed network-based malicious URL identification algorithm.

\par
\begin{figure*}[htb]
\centering
\includegraphics[width=\textwidth]{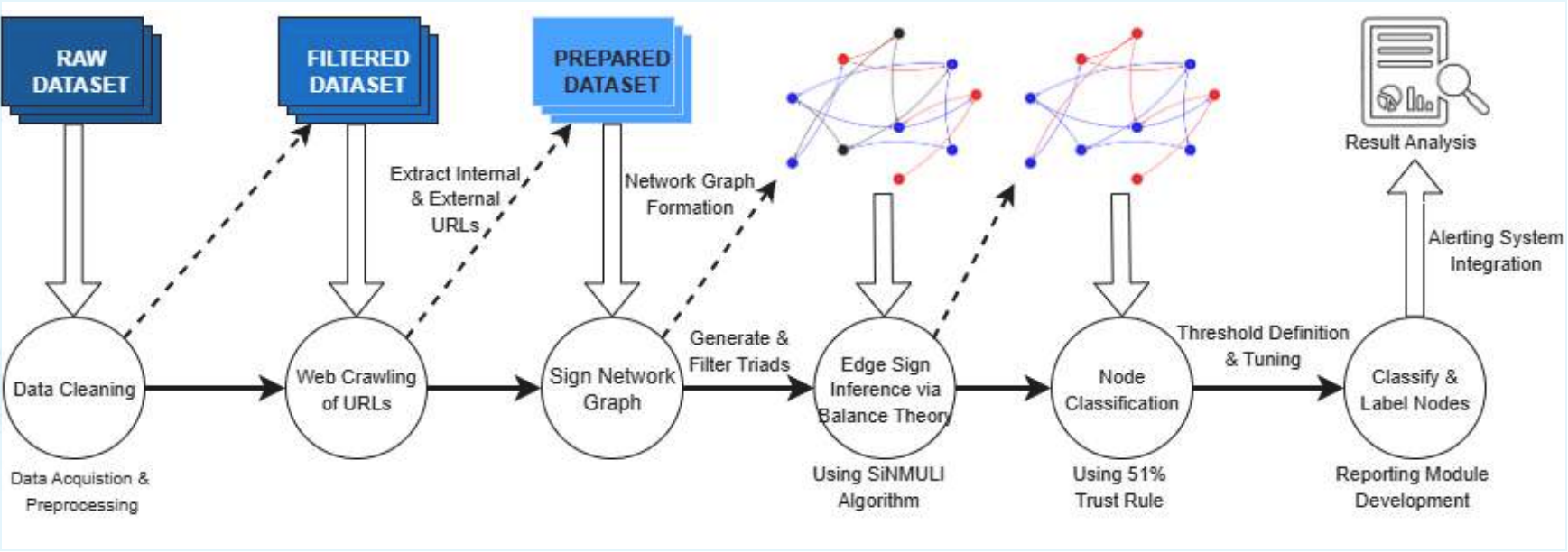}
\caption{The figure represents the schematic diagram of the proposed SiNMULI workflow for the identification of malicious URLs using a signed network approach.}
\label{fig:SD}
\end{figure*}

 The proposed \emph{SiNMULI} framework employs a structured pipeline for malicious website detection based on signed network analysis as described in the given schematic diagram in figure~\ref{fig:SD}. Beginning with a benchmark dataset, the system first extracts both internal and external URLs, where each unique domain or subdomain is represented as a distinct URL stored further for hyperlink signed network formation. From the developed signed network, we identify the triadic relationship among the URLs and filter to form the fundamental units for structural analysis. Leveraging our proposed directed balance theory, the algorithm infers missing edge signs to maintain structural consistency across triads. Subsequently, the label of the unknown URLs is determined using the \emph{51\% majority rule}, wherein a node is labeled according to the predominant sign of its incident edges. This integrated inference and classification process enables reliable categorization of an unlabeled URL as benign or malicious. In the final stage, classification results are aggregated to produce actionable outputs for downstream security analysis and threat mitigation.

\subsection{Seed Dataset}\label{s:dataset}In our work, we have used a labeled benchmark dataset for collecting the hyperlink relation among the URLs. We employ the publicly available \emph{Phishing Websites Dataset}~\cite{phishing_dataset} (Ariyadasa \textit{et al.}) as a seed dataset. It is a large-scale corpus of legitimate and phishing webpages. The rationale behind the selection of this benchmark dataset as a seed is its large chunk of data points, as well as being labeled either as legitimate or phishing. The dataset is also very lightweight due to the minimal information regarding each URL compared to other existing benchmark datasets.

Legitimate pages were sourced from two primary channels: (i) the top 5 results of keyword-based Google searches with domain diversity constraints (maximum of 10 URLs per domain), and (ii) the Ebbu2017 dataset, contributing approximately 25,874 active legitimate URLs. Phishing pages were obtained from three well-established threat intelligence repositories: \textit{PhishTank} (December 2020 -- October 2021), \textit{OpenPhish} (September -- October 2021), and \textit{PhishRepo} (September -- October 2021). Automated scripts continuously monitored these feeds to capture live phishing pages, ensuring minimal loss due to the ephemeral nature of such pages.

\subsubsection{Summary of Seed dataset} 


\begin{table*}[t]
    \centering
    \caption{Seed dataset summary and working 3-class distribution in the seed index}
    \label{tab:seed-summary-classes}  

    \begin{subtable}{0.60\textwidth}
        \centering
        \caption{Seed dataset summary}
        \label{tab:seed-summary}

        \begin{tabularx}{\linewidth}{l X}
            \hline
            \textbf{Metric} & \textbf{Value} \\
            \hline
            Source benchmark & Phishing Websites Dataset (Ariyadasa et al.) \\
            Raw URLs (before cleaning) & 83{,}275 \\
            Retained URLs (after cleaning) & 80{,}000 \\
            Label mapping (working index) & $+1$ legitimate, $-1$ malicious, $0$ unknown \\
            Exact duplicates removed & 121 \\
            Columns (schema) & \texttt{URL} (string), \texttt{Label} (int), \texttt{source} (string) \\
            Storage format & CSV (UTF-8), one URL per row \\
            \hline
        \end{tabularx}
    \end{subtable}
    \hfill
    \begin{subtable}{0.35\textwidth}
        \centering
        \caption{Working 3-class distribution in the seed index}
        \label{tab:seed-classes}

        \begin{tabular}{lrr}
            \hline
            \textbf{Class} & \textbf{Count} & \textbf{Percent} \\
            \hline
            Legitimate ($+1$) & 45{,}000 & 56.25\% \\
            Malicious ($-1$) & 30{,}000 & 37.50\% \\
            Unknown ($0$) & 5{,}000 & 6.25\% \\
            \hline
            \textbf{Total} & 80{,}000 & 100\% \\
            \hline
        \end{tabular}
    \end{subtable}

\end{table*}

Table~\ref{tab:seed-summary} reports the core properties of the seed. For transparency, we also report the class distribution of the working three-class index (used for all downstream analyses and ablations) in table~\ref{tab:seed-classes}. Percentages are with respect to the 80{,}000 retained rows.


\par{\bf \em Usage downstream.}
The seed is \emph{not} the signed graph itself; it is the anchor list for systematic crawling and hyperlink extraction. After crawling, we canonicalize hosts, type internal/external links, and construct a directed signed network (nodes = sites/domains; edges = hyperlinks with trust/distrust semantics). Missing edge signs are inferred via balance-theoretic constraints on triads, and unlabeled nodes are resolved using a simple majority rule on incident edge signs. The resolved labels feed the subsequent evaluation and analysis.
\subsection{Dataset Description}

The dataset used in this study, \texttt{mysql\_index\_dataset.csv}, is the starting point for building the signed hyperlink network analysed in the SiNMULI framework. It contains indexed web domains labelled as \emph{legitimate (B)}, \emph{malicious (M)}, or \emph{unknown/unlabeled (U)}. Each row represents a single website (domain) and records what we learned from large-scale crawls: basic host information and the outgoing links it makes to other sites. The resulting directed graph has \(48{,}205\) nodes and \(384{,}592\) edges, with a directed density of about \(1.655\times 10^{-4}\), so it is large but very sparse. Most nodes are already labelled---\(46{,}415\) benign (B) and \(1{,}790\) malicious (M)---while the remaining unlabeled nodes are used for semi-supervised learning and triad-based balance checks. The graph naturally splits into \(362\) communities; the largest contains roughly \(19{,}062\) nodes (\(18{,}453\) B and \(609\) M), showing that large benign clusters and smaller adversarial pockets coexist in the same network. We also identify \(2{,}236\) \emph{articulation points} and \(4{,}566\) \emph{bridge edges}, whose removal would disconnect parts of the graph. These weak points are important for understanding both network resilience and how signals propagate during inference.

From a research perspective, this dataset captures the inherent properties of a real-world signed network where trust and hostility co-evolve. The high occurrence of legitimate nodes indicates a trust-dominated structure, while the dense existence of malicious edges brings about polarity and weak balance conditions within communities. The articulation points and bridges indicate regions sensitive to topological disruption, making them ideal for studying propagation decay, infiltration, and defensive graph rewiring strategies. Overall, this dataset supports multi-level analyses --- from local triadic inference to global polarity stability --- and provides a robust empirical foundation for validating the SiNMULI approach in cybersecurity-aware signed graph learning.

\subsection{Data Collection}\label{s:datacollectionAlgo} In this section, we describe the data collection process based on the benchmark seed dataset. We describe the web crawling from the seed dataset and further filtering and preprocessing of the crawled URLs for the development of a signed hyperlink network. We also propose a method for data collection in a formal way, using an algorithmic approach, and further analyze this algorithm to reveal the computational complexity. This formal presentation of the algorithm helps in the reproducibility of the dataset\footnote{All code and dataset used in this study---including data loaders, training scripts, pre-trained checkpoints, and the code that generates every figure and table in this manuscript for analysis/visualization pipelines---is openly available at: \url{https://github.com/sayanmondal2098/SinMuli}}. The proposed algorithm has mainly three segments of task: a) Web scraping, b) Cleaning \& Preprocessing, c) Output preparation. In the subsequent section, we describe the phases in detail, one after another.

\subsubsection{Web scraping} In this step, we perform a one-hop crawl of external hyperlinks discovered during the scraping process. Specifically, every external link appearing on the seed pages was treated as a new candidate URL. Each of these external URLs was then fetched once, and their content and metadata were collected in the same manner as the seed dataset. This one-hop strategy lets us capture broader connections in the malicious ecosystem while keeping the dataset size manageable. The added external links brought in more malicious and legitimate pages and, at the same time, exposed useful structural and relational patterns between sites.

Web scraping also comes with practical constraints, and we design around them. To reduce the chance of being blocked, we throttle requests, randomise intervals, and back off when error rates rise. When pages depend on JavaScript, we first try to use server-rendered HTML and switch to a headless browser only when needed. CAPTCHAs and strict bot checks are respected: if they appear, we stop collection rather than attempt to bypass them. Because real-world markup is often messy, our parser is fault-tolerant and can fall back to alternative strategies. To cut down link noise, we normalise and de-duplicate URLs, follow redirects, and store both the requested and final targets. To limit seed bias, we draw from multiple feeds and periodically review samples to maintain data quality.

\subsubsection{Data Cleaning \& Preprocessing}
The structure and density of the final signed network depend directly on how the data is cleaned and preprocessed. By removing structurally weak nodes, we increase the number of triads, reduce sparsity, and improve the reliability of downstream algorithms such as malicious URL classification and sign inference. The cleaning stage ensures that: (a) sparse or dangling nodes are removed; (b) dense, triad-rich subgraphs needed for balance-theoretic reasoning are preserved; and (c) classification of malicious URLs becomes more stable and accurate.

To keep the constructed signed network structurally sound and analytically consistent, we apply a multi-stage cleaning procedure as described below.

\paragraph{Structural Validation of URLs}Before building the signed network, we first check whether each URL has meaningful hyperlink connections. In particular, we verify the presence of at least one internal or external link, which is required for triadic inference. URLs without such links are discarded to maintain a well-connected and informative graph. URLs that don't have these links are removed to keep the structure strong and accurate. Let \( u \in U_0 \) denote a URL from the seed dataset. A URL \( u \) is retained only if: $|L_{\text{int}}(u)| + |L_{\text{ext}}(u)| > 0$, where \( L_{\text{int}}(u) \) and \( L_{\text{ext}}(u) \) denote the sets of internal and external hyperlinks, respectively.

\paragraph{Noise Reduction}Real-world web data often contains noise in the form of broken links, unreachable pages, or malformed content. In this stage, we remove such problematic entries so that weakly connected or fragmented components do not distort the analysis. This produces a cleaner, more coherent graph and improves the reliability of subsequent network-based inference.

\paragraph{Output Graph Consistency}
To create a clear and easy-to-analyse signed network, we only finalise the nodes and connections after checking and cleaning the data: we standardise the hosts and combine all URLs from the same host into one node, remove broken or self-linking, JavaScript-based, tracking, and duplicate links, and fix any redirect issues. Each remaining hyperlink is typed as \emph{internal} (same host) or \emph{external} (different host) and assigned a sign that preserves trust/distrust semantics: \(+1\) for edges originating from labeled legitimate sources and \(-1\) for edges originating from labeled malicious sources; edges from unlabeled sources are retained but left unsigned.

\subsubsection{Output preparation}After the data cleaning and preprocessing of the collected data, we only keep those meaningful URLs in the final dataset, which can further contribute to our signed hyperlink network formation. In this process, many URLs lack meaningful structural data, and we filter out pages without usable hyperlink structure or those returning inaccessible content (e.g., 403/404). Thus, only structurally valid URLs are kept in our final dataset to participate in network formation that improves connectivity and reliability. Further, remaining URLs and links are consolidated into a compact JSON artefact, preserving node features for signed-network construction and classification. 
\par We formally describe our proposed algorithm~\ref{algo:datacollection} for data collection, and afterwards we critically analyze the data collection algorithm to understand the computational complexity.










\begin{algorithm}[H]
\caption{Data Collection based on Benchmark dataset}
\label{algo:datacollection}
\KwIn{Seed URL list $\mathcal{U}_0$}
\KwOut{Filtered dataset $\mathcal{J}$ for signed network analysis}
$V \leftarrow \mathcal{U}_0$\; \tcp*{Initialize node set from URLs}
$E \leftarrow \emptyset$\; \tcp*{Initialize edge set for hyperlink structure}
\ForEach{$u \in V$}{
  $c(u) \leftarrow$ Download and parse HTML content\;
  $L_{\text{int}}(u) \leftarrow$ Extract internal links\;
  $L_{\text{ext}}(u) \leftarrow$ Extract external links\;
  \If{$|L_{\text{int}}(u)| = 0$ \textbf{and} $|L_{\text{ext}}(u)| = 0$}{
    Remove $u$ from $V$\; \tcp*{No network structure: discard}
    \textbf{continue}\;
  }
  $s(u) \leftarrow$ Check SSL status (HTTPS or HTTP)\;
  $r(u) \leftarrow$ Query reputation service (e.g., SafeBrowsing, optional)\;
  $J(u) \leftarrow \{ L_{\text{int}}(u), L_{\text{ext}}(u), s(u), r(u) \}$\;
  Add $L_{\text{int}}(u) \cup L_{\text{ext}}(u)$ to $E$\;
}
$\mathcal{J} \leftarrow \{ J(u)\ \forall u \in V \}$\;
\KwRet{$\mathcal{J}$ for downstream signed network analysis}\;
\end{algorithm}

\par{\bf \em Analysis of Proposed Data collection Algorithm:} This section reveals the computation complexity of the proposed data collection algorithm. To analyze the proposed data collection algorithm, we have systematically segregated the tasks performed in the algorithm and individually computed the complexity of each segment of the task.
\paragraph{Web Scraping}
To build the hyperlink graph for network modelling, we first scrape and parse the HTML for each URL in the seed set. This step records internal/external links, SSL status, and basic metadata, producing a structurally rich view of each site. Pages are processed independently for scalability. For each \(u \in \mathcal{U}_0\):

\begin{itemize}
  \item HTML parsing runs in \(\mathcal{O}(L)\), where \(L\) is the page size.
  \item Anchor scanning (internal/external link extraction) is \(\mathcal{O}(L)\) with post-processing/storage \(\mathcal{O}(m)\), where \(m\) is the number of discovered hyperlinks.
\end{itemize}

Let \(n = |\mathcal{U}_0|\) and \(m\) denote the average hyperlinks per page. The total time for scraping and link parsing is
\[
\mathcal{O}\bigl(n\,(L+m)\bigr).
\]

\paragraph{Filtering and URL Extraction}
Many URLs contribute little structural signal. We therefore drop pages with no link structure or inaccessible content (e.g., 403/404). This improves graph connectivity and downstream reliability. Per-URL costs are:
\begin{itemize}
  \item SSL status check: \(\mathcal{O}(1)\).
  \item Optional reputation lookup (cached/parallelised): \(\mathcal{O}(1)\) amortised.
  \item Parsing and storing hyperlink features: \(\mathcal{O}(m)\).
  \item \textbf{Feature extraction per valid URL:} \(\mathcal{O}(m)\).
\end{itemize}

\paragraph{Output Preparation}
After filtering, we consolidate valid pages \(V' \subseteq V\) into a structured JSON suitable for signed-graph construction, preserving node-level features:
\[
|V'| \le n, \qquad \text{total output space } \mathcal{O}(n' m),\ \text{where } n' = |V'|.
\]

\paragraph{Overall Complexity}
The collection and parsing stage costs \(\mathcal{O}\bigl(n\,(L+m)\bigr)\) time---covering seed URL scraping and a single hop over discovered out-links---while writing the curated JSON uses \(\mathcal{O}(n' m)\) space. In practice, the pipeline parallelises naturally over seeds: independent asynchronous crawlers enforce per-domain rate limits, cache redirects/reputation results, invoke headless rendering only when needed, and aggressively prune non-structural pages to shrink \(|E|\), reducing compute and I/O overhead.

\subsection{Summary of Collected Dataset}In this section, we analyze the collected dataset and find the insights of the dataset that strengthen our understanding of the data, which helps us to develop the signed hyperlink network further. From our parsed dataset, we identified a total of 261,113 URLs. Among them, 79,759 URLs are labeled using the provided ground truth, comprising 49,868 legitimate URLs and 29,891 malicious URLs. The remaining 181,354 URLs are unlabeled, typically discovered as external links or indirect connections during HTML parsing. In total, 684,114 directed hyperlinks were extracted from the hyperlink structure, forming the backbone of the directed graph. Notably, 685,552 of these are external edges, where the destination lies outside the original labeled dataset. These external connections are crucial in modeling real-world web browsing behavior, where benign and malicious pages often link to or are linked from external domains. 

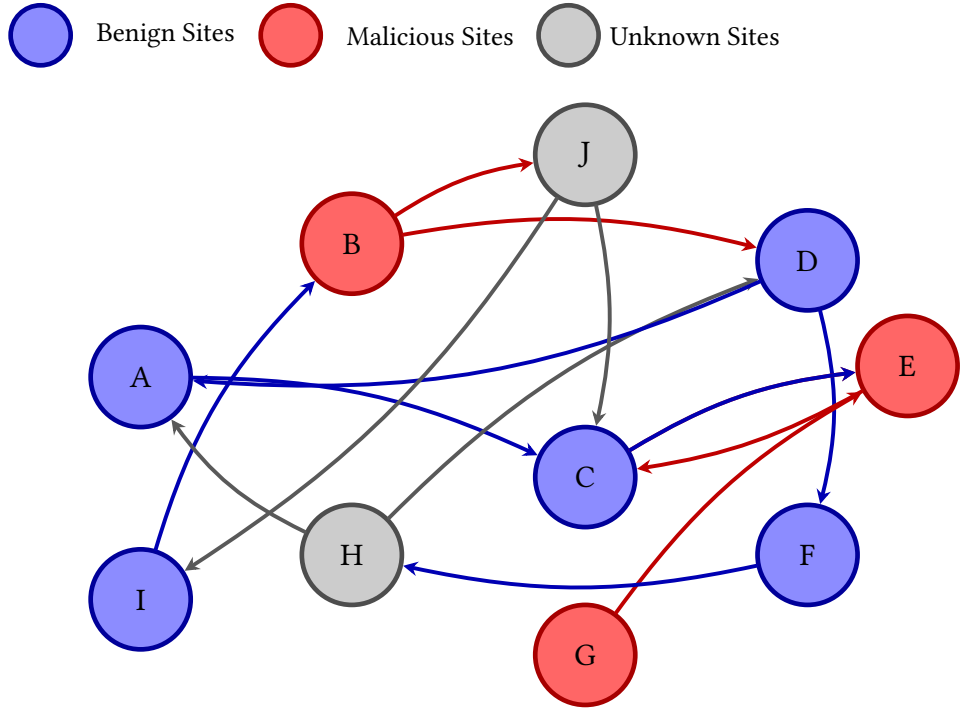
\begin{figure}[t]
\centering
\resizebox{0.80\linewidth}{!}{
\begin{tikzpicture}[>=Stealth, line cap=round, line join=round]

\tikzset{
  v/.style           ={circle, minimum size=9mm, very thick, font=\footnotesize},
  legitnode/.style   ={v, draw=blue!60!black,  fill=blue!45},
  phishnode/.style   ={v, draw=red!65!black,   fill=red!60},
  unknownnode/.style ={v, draw=gray!60!black,  fill=gray!40},
  extlegit/.style    ={draw=blue!70!black,  line width=1.0pt, -{Stealth[length=1.3mm,width=1.3mm]}},
  extphish/.style    ={draw=red!75!black,   line width=1.0pt, -{Stealth[length=1.3mm,width=1.3mm]}},
  extunknown/.style  ={draw=gray!70!black,  line width=1.0pt, -{Stealth[length=1.3mm,width=1.3mm]}},
  selfloop/.style    ={->, draw=black, line width=1.0pt},
}

\node[legitnode,   minimum size=5.4mm] (legdot) at (-3.90, 2.325) {};
\node[anchor=west, font=\scriptsize]                 at (-3.52, 2.325) {Benign Sites};

\node[phishnode,   minimum size=5.4mm] (phdot)  at (-1.65, 2.325) {};
\node[anchor=west, font=\scriptsize]                 at (-1.27, 2.325) {Malicious Sites};

\node[unknownnode, minimum size=5.4mm] (undot)  at (0.85, 2.325) {};
\node[anchor=west, font=\scriptsize]                 at (1.10, 2.325) {Unknown Sites};

\begin{scope}[shift={(0,-0.75)}]  

\node[legitnode]   (A) at (-3.0,  0.0)  {A};
\node[phishnode]   (B) at (-1.1,  1.2)  {B};
\node[legitnode]   (C) at ( 1.0, -0.9)  {C};
\node[legitnode]   (D) at ( 3.0,  1.05) {D};
\node[phishnode]   (E) at ( 3.9,  0.1)  {E};
\node[legitnode]   (F) at ( 3.0, -1.6)  {F};
\node[phishnode]   (G) at ( 1.0, -2.5)  {G};
\node[unknownnode] (H) at (-1.1, -1.6)  {H};
\node[legitnode]   (I) at (-3.0, -2.0)  {I};
\node[unknownnode] (J) at ( 1.0,  2.0)  {J};

\draw[extlegit,   bend left=12] (A) to (C);
\draw[extlegit,   bend left=14] (D) to (F);
\draw[extlegit,   bend left=14] (I) to (B);

\draw[extphish,   bend left=12] (B) to (D);
\draw[extphish,   bend left=12] (B) to (J);
\draw[extphish,   bend left=12] (C) to (E); 
\draw[extlegit,   bend left=12] (C) to (E);
\draw[extphish,   bend left=12] (G) to (E);
\draw[extphish,   bend left=10] (E) to (C);

\draw[extunknown, bend left=14] (H) to (A);
\draw[extunknown, bend left=12] (H) to (D);
\draw[extlegit,   bend left=12] (F) to (H); 
\draw[extlegit,   bend left=14] (D) to (A);
\draw[extunknown, bend left=12] (J) to (C);
\draw[extunknown, bend left=12] (J) to (I);


\end{scope}
\end{tikzpicture}
}
\caption{Signed hyperlink network structure showing legitimate (blue) and phishing (red) sites and the unlabeled URLs(grey). The color of the edges corresponds to the color of the originating node.}
\label{fig:network}
\end{figure}

\subsection{Network Formation}In this section, we describe the formation of a signed hyperlink network. We describe the systematic formation of a network. Further, we analyze the network to understand the distinct behavior of the malicious URLs from legitimate URLs. 
\par We represent the URLs as nodes, and the hyperlinks among them have been represented as edges among the nodes. In general, the URLs in the cyber world can be categorized into two basic types: a) main Website URL, b) subdomain URL. Thus, the hyperlinks formed between the URLs have also been classified as an internal edge: It connects between the main website URL and the subdomain URL. b) external edge: It connects different main website URLs. As the subdomains belong to the main website, we can simply consider the internal edges as a self-loop, which represents the existence of internal edges among the main website and its subdomains. However, in our proposed work, as the back-links play a major role at the time of network formation, we have considered all subdomain URLs of a main URL as a single node representation. The composition of multiple URLs of the same domain as a single node does not lose the role of a hyperlink network in our proposed model. To better understand the hyperlink structure of benign and malicious websites, we visualize the signed network by a schematic diagram as shown in Figure~\ref{fig:network}. In this schematic diagram, nodes represent domains, while directed edges indicate external hyperlinks among themselves. In our network, we have specifically three different kinds of nodes, i.e., a)malicious, b)benign, and c)unlabeled. The signs on the edges in the network are marked as indicated in the collected dataset. We have kept the back-links originating from malicious nodes as $(-1)$ and $(+1)$ for the edges originating from benign nodes. However, the edges from the unlabeled nodes are left unsigned at the initial phase of network formation.

\par{\bf \em Network Analysis:}

Our seed is a list of \emph{URLs} (80{,}000 rows), while the interaction network is constructed over \emph{domains/hosts} that actually participate in hyperlink interactions during crawling. The reduction arises from: (i) URL$\!\to\!$domain canonicalization (many URLs collapse to a single host node), (ii) removal of degree-$0$ nodes (no observed links), (iii) scope/media constraints (robots, login gates, non-HTML, HTTP errors), (iv) deduplication of repeated URL and URL+hop rows, and (v) partial label coverage (only hosts overlapping the seed can be label-stratified). Together, these steps yield a directed graph with markedly fewer nodes than the raw URL count.

We summarize the detailed network analysis in table~\ref{tab:two_tables}. We analyzed a directed interaction graph with \(N=48{,}205\) nodes and an estimated \(E=384{,}592\) edges (computed as \(\sum \text{outdeg}\)). The network is sparse, with density \(\rho = \tfrac{E}{N(N-1)} \approx 1.66\times 10^{-4}\). Degree statistics indicate skew: the mean out-degree is \(7.98\) while the median is \(5\); the median in-degree is \(1\). Community detection yields \(362\) modules; the largest comprises \(19{,}062\) nodes, of which \(96.8\%\) are labeled benign (B) and \(3.2\%\) malicious (M). Structural vulnerability is non-trivial: we identify \(2{,}236\) articulation points and \(4{,}566\) bridge edges, implying single-edge/node failure can disconnect portions of the graph. Labels are imbalanced overall (B \(=46{,}415\), \(96.29\%\); M \(=1{,}790\), \(3.71\%\)).

Label-stratified centrality shows systematic separation between benign and malicious nodes. Benign nodes occupy more central and connective positions: median out-degree \(5\) (vs.\ \(2\) for M) and median in-degree \(1\) (vs.\ \(0\) for M). Eigenvector centrality, which captures recursive influence, is higher for benign nodes (\(\tilde{e}_{B}=2.538\times 10^{-3}\) vs.\ \(\tilde{e}_{M}=1.336\times 10^{-4}\)). PageRank medians are close in magnitude but still lower for malicious nodes (\(\tilde{p}_{B}=4.435\times 10^{-6}\) vs.\ \(\tilde{p}_{M}=3.720\times 10^{-6}\)). Median betweenness centrality is \(0\) for both labels, consistent with a sparse, modular graph in which most vertices do not lie on many shortest paths. Taken together, these observations suggest malicious actors tend to be peripheral and poorly embedded within dominant communication backbones.

\begin{table*}[t]
\centering
\caption{Macro-level structural summary and label-stratified node centrality statistics of the signed network.}
\label{tab:two_tables}

\begin{minipage}[t]{0.47\textwidth}
\centering
\textbf{(a) Macro (graph-level) summary}
\begin{tabularx}{\linewidth}{l r}
\hline
\textbf{Property} & \textbf{Value} \\
\hline
Nodes ($N$) & \makecell{48{,}205\\ legit.(46{,}415)\\Mal.(1{,}790)} \\
Directed edges ($E$) & 384{,}592 \\
Density $\rho=\frac{E}{N(N-1)}$ & $1.655 \times 10^{-4}$ \\
Communities (count) & 362 \\
Largest community (nodes) & \makecell{19{,}062\\ legit.(18{,}453)\\Mal.(609)} \\
Largest community label mix &
\makecell[l]{96.8\% Legitimate\\3.2\% Malicious} \\
Articulation points & 2{,}236 \\
Bridge edges & 4{,}566 \\
\hline
\end{tabularx}
\end{minipage}
\hfill
\begin{minipage}[t]{0.47\textwidth}
\centering
\textbf{(b) Label-stratified node centrality (medians)}
\begin{tabular}{|p{2.8cm}|p{1.2cm}|p{1.2cm}|p{1.2cm}|}
\hline
\textbf{Centrality} & \textbf{Legit.} & \textbf{Mal.} & \textbf{Gen.} \\
\hline
Out-degree                 & 5 & 2 & 5 \\
In-degree                  & 1 & 0 & 1 \\
Eigenvector $\tilde{e}$   & $2.54\times10^{-3}$ & $1.34\times10^{-4}$ & $2.50\times10^{-3}$ \\
PageRank $\tilde{p}$      & $4.44\times10^{-6}$ & $3.72\times10^{-6}$ & $4.37\times10^{-6}$ \\
Betweenness $\tilde{C}_B$ & $4.90\times10^{-8}$ & $1.62\times10^{-11}$ & $0.00$ \\
Closeness $\tilde{C}_H$   & $100308.3$ & $82798.1$ & $78266.5$ \\
Clustering $\tilde{C}$    & $0.2909$ & $0.3000$ & $0.2693$ \\
\hline
\end{tabular}
\end{minipage}

\end{table*}

\subsection{Proposed Malicious URL detection Algorithm: S{\em i}NMULI} In this section, we formally propose the signed hyperlink network-based malicious URL identification algorithm, S{\em i}NMULI. Our proposed algorithm is based on balancing the triadic relationships in the signed hyperlink network. In this algorithm, we have three distinct parts: a) {\em Triad identification}, b) {\em Sign determination of unlabeled edges in the triad}, c) {\em Node labeling using majority voting}. 
\paragraph{Triad Identification}Initially, we extract all 3-node subgraphs (triads) \( T \subseteq V^3 \) from the signed network. This triadic relationship plays a vital role in further identification of the nature of unlabeled nodes. We specifically keep only those triads for our further observation for balancing, which contain one and only one unlabeled node with an unlabeled edge. Other kinds of triads are filtered out from our study. The objective of this filtering is to keep only those triads where unlabeled URLs have only one back-link in that concerned triad, which can be further inferred using the proposed directed balance theory. We find that many triads share common edges among themselves. This overlapping nature further helps to determine the sign of the unlabeled edges in the network, and thereby determines the label of the node further.

 \paragraph{Edge Sign Determination}We further determine the sign of the unlabeled edge in a given triad \(T\), which consists of nodes \( (u, v, w) \), where exactly one node is unlabeled. We apply the proposed directed balance theory in the signed hyperlink triad to determine the sign of the unlabeled edge. In this method, we consider either weak or strong balance for all the possible cases of balancing situation described in figure~\ref{fig:direct_bal-cases}.

 \paragraph{Node Labeling}Finally, we classify the unlabeled nodes into benign or malicious in the network with the help of the inferred edge sign. As there exist many overlapping triads in the signed hyperlink network, we apply $51\%$ majority voting rule to identify the final label of the unlabeled edges in the network. Based on the final edge label. We determine the nature of the unlabeled node. We describe the algorithm formally in the following as algorithm~\ref{algo:sinumuli}.

\par{\bf \em Analysis of Proposed S{\em i}MULI Algorithm: }In this section, we analyze the proposed S{\em i}NMULI algorithm to measure the requirement of computational resources to run this framework to identify the malicious URLs. This formal method of analysis also helps us to compare the effectiveness of the proposed model over the existing baselines. As the proposed algorithm has three distinct sections, we also analyze every individual section carefully and find the overall complexity of the proposed model.  

\par Let the signed directed network be denoted by \(G(V,E)\), with \(V\) being the set of nodes (websites) and \(E\) the set of edges (hyperlinks), each edge \(e_{uv}\) carrying a sign \(s_{uv}\in\{+1,-1\}\). The maximum possible number of triads in the network would be \(\binom{|E|}{3}\), and assume \(T\) is the set of triads kept after filtering, thus \(T \subseteq \binom{|E|}{3}\). Therefore the complexity of triad identification can be expressed as \(\mathcal{O}\!\left(\binom{|E|}{3}\right)\), i.e., \(\mathcal{O}(|E|^3)\). 

The inference of the sign of the unlabeled edge using the proposed directed balance theory involves iteratively assigning missing edge signs to ensure this triadic balance. Given \(|T|\) total triads in the graph, the worst-case complexity of triadic edge inference per iteration is \(\mathcal{O}(|T|)\), therefore it can be expressed as \(\mathcal{O}(|E|)\) as \(T \subseteq \binom{|E|}{3}\). Finally, after edge signs are inferred, unlabeled nodes \(v\in V\) are classified based on the $51\%$ majority voting rule. As each unlabeled node of set \(v\) belongs to the subset of \(T\). Thus, the computational complexity of labelling is \(\mathcal{O}(|v|)\). Thus it can be deduced to \(\mathcal{O}(|V| )\). Therefore, overall computational complexity could be estimated by combining the three phases of the proposed algorithm as \(\mathcal{O}(|V| + |E|)\). On the other hand, the memory (space) complexity for maintaining the network representation (adjacency structure and node labels) is \(\mathcal{O}(|V| + |E|)\).

\par The algorithm leverages structural balance to deliver robust and interpretable malicious URL detection. The iterative inference method efficiently converges due to typically sparse real-world hyperlink structures, and the node labeling step provides an intuitive mapping between network trust dynamics and classification outcomes, enhancing both explainability and detection accuracy.

\section{Result \& Discussion}\label{s:result}
In this section, we empirically evaluate the proposed S{\em i}NMULI framework on the signed hyperlink network constructed from our crawled dataset. We first describe how ground-truth labels are established using independent threat-intelligence sources and then visualize representative local signed subgraphs to build intuition about how benign and malicious sites are embedded in the network. Next, we characterize the distribution of the seven triad types defined in Section~IV and relate their prevalence to strong and weak balance configurations. Finally, we present quantitative detection results, analyze standard evaluation metrics, and compare S{\em i}NMULI against state-of-the-art blacklist, traditional machine-learning, deep-learning, and graph-based baselines.

\subsection{Ground Truth for Validation} To validate the effectiveness of our proposed model, we establish the ground truth labels through independent verification using authoritative third-party threat intelligence services. Specifically, the unlabeled URLs of the test set are cross-validated against public reputation databases, PhishTank, and OpenPhish. These platforms maintain updated blacklists of confirmed phishing URLs contributed and vetted by the security community. This external lookup allows us to independently confirm the correctness of our model's predictions without relying solely on internally annotated labels. We only include URLs that receive an exact match from the third-party database to ensure high-confidence validation. This external validation methodology enhances the reliability of our evaluation framework by introducing an unbiased reference standard. The high agreement observed between our predictions and the third-party ground truth supports the trustworthiness and real-world applicability of the S{\em i}NMULI framework.
\begin{algorithm}[hbt]
\caption{Signed Network Edge Sign Inference and Node Classification}
\label{algo:sinumuli}
\KwIn{Filtered dataset $\mathcal{J}$ (from Algorithm~\ref{algo:datacollection})}
\KwOut{Node label set $\mathcal{Y} \in \{+1,-1\}$ (legitimate or malicious)}
Initialize signed directed graph $G(V,E)$ from dataset $\mathcal{J}$\;
Set initial node labels: legitimate $(+1)$, malicious $(-1)$, unknown $(0)$\;
Set initial edge signs $s_{uv}$: legitimate $\rightarrow$ legitimate $(+1)$, involving malicious $(-1)$, unknown $(0)$\;
\tcp{Triad identification}
$T \leftarrow \mathsf{AllDirectedTriads}(G)$\;
$T^\star \leftarrow \{\tau\in T:\mathsf{countUnknownEdges}(\tau)=1 \wedge \mathsf{countUnlabeledNodes}(\tau)=1\}$\;
Build index $\mathsf{TriadsByEdge}[(u,v)] \mapsto \{\tau\in T^\star:(u,v)\in\tau\}$ and set $Q\leftarrow T^\star$\;
\tcp{Edge sign inference via balance theory}
\ForEach{triad $(u,v,w)\in Q$}{
  \If{exactly one unlabeled edge exists}{
    Let $e_{?}=x\!\to\!y$\; \tcp*{$e_{?}$: unlabeled edge}
    If $y(x)=+1$, prefer $s_{xy}=+1$; if $y(x)=-1$, prefer $s_{xy}=-1$\;
    Infer the edge sign to satisfy strong balance, otherwise weak balance\;
    Update $s(e_{?})$ and enqueue $\tau'\in\mathsf{TriadsByEdge}[e_{?}]$\;
  }
}
\tcp{Node labeling via majority voting}
\ForEach{unlabeled node $v\in V$}{
  Let $E_v^-=\{(u,v)\in E:s_{uv}=-1\}$ and $E_v^+=\{(u,v)\in E:s_{uv}=+1\}$\;
  If $|E_v^+|+|E_v^-|=0$, continue; set $\textsc{Majority}=0.51$ and $\textsc{TiePolicy}\in\{\text{abstain},\text{malicious},\text{benign}\}$\;
  Assign $y_v$ according to $\displaystyle y_v=\begin{cases}
  +1, & \frac{|E_v^+|}{|E_v^+|+|E_v^-|}>\textsc{Majority},\\
  -1, & \frac{|E_v^-|}{|E_v^+|+|E_v^-|}>\textsc{Majority},\\
  0/-1/+1, & \text{otherwise (by \textsc{TiePolicy}).}
  \end{cases}$\;
}
\KwRet{Final node labels $y_v$}\;
\end{algorithm}

\subsection{Visualising Local Signed Subgraphs}

\begin{figure}[t]
  \centering
  \subfloat[Before inference\label{fig:cross-main-before}]{
    \includegraphics[width=0.47\columnwidth]{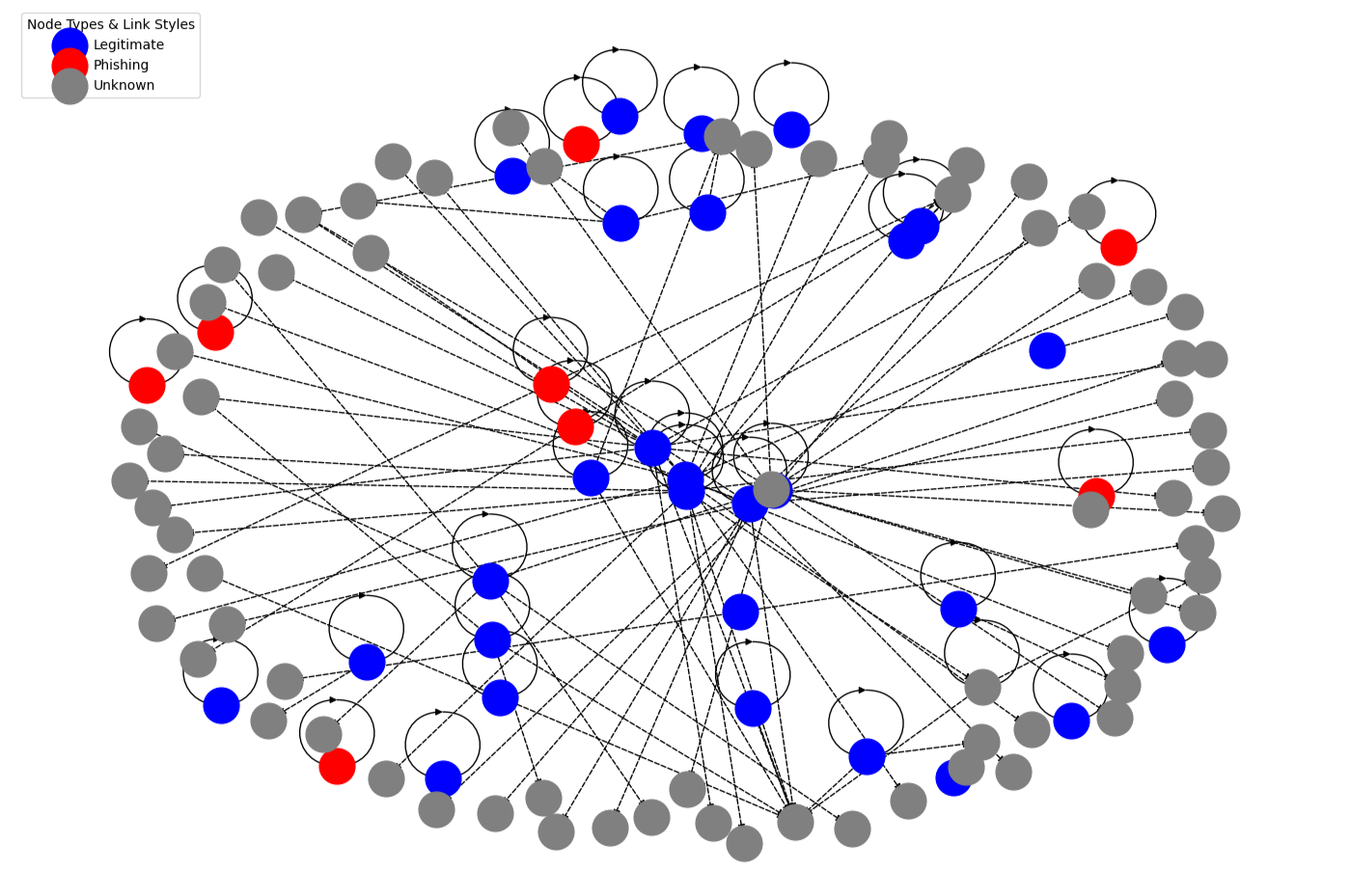}%
  }\hfill
  \subfloat[After S{\em i}NMULI inference\label{fig:cross-main-after}]{
    \includegraphics[width=0.47\columnwidth]{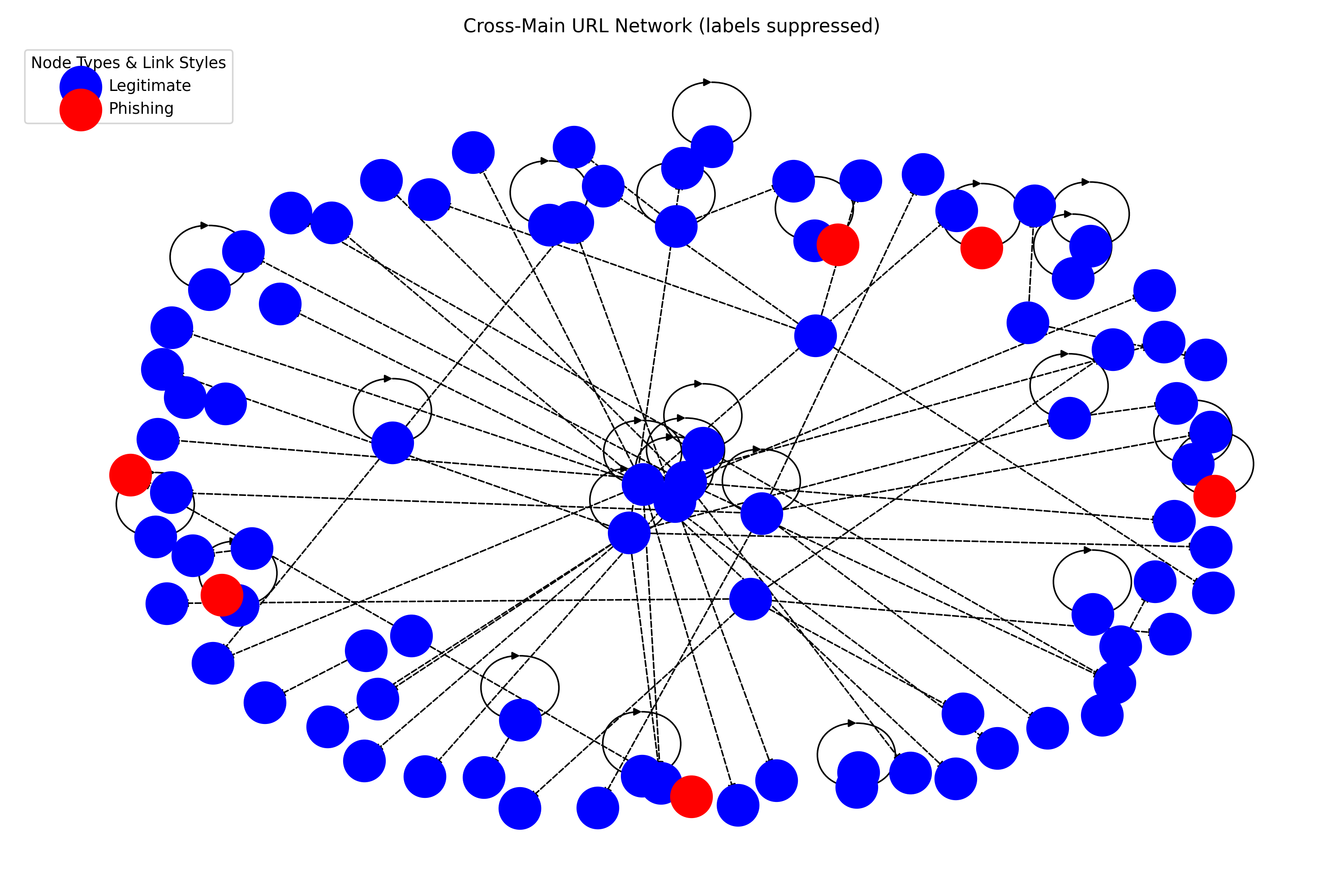}%
  }
  \caption{Figure shows the example signed hyperlink subgraph containing a small set of URLs among which a few are malicious, and some are legitimate, whereas the majority are unlabeled. Blue nodes denote legitimate URLs, red nodes denote malicious URLs, and grey nodes are unlabeled. The figure~\ref{fig:cross-main-before} corresponds to the network structure prior to the identification of unlabeled URLs by SiNMULI. Whereas, the figure~\ref{fig:cross-main-after} corresponds to the network structure after the identification of unlabeled URLs.}
  \label{fig:cross-main-before-after}
\end{figure}

To build intuition for how S{\em i}NMULI propagates trust signals, we first inspect a small local region of the signed network centered on a popular cross-domain infrastructure site. Such hubs are important because they receive links from many heterogeneous domains and can act as conduits through which both benign and malicious influence spreads. By examining a concrete example, Fig.~\ref{fig:cross-main-before-after}, we can see how balance-based inference reshapes local neighbourhoods around these shared infrastructure nodes. To qualitatively inspect how phishing URLs are embedded within the hyperlink structure, we examine a representative subgraph centered on a popular cross-domain infrastructure site. Fig.~\ref{fig:cross-main-before-after} shows the same subgraph before and after S{\em i}NMULI's balance-based inference.

Before inference (Fig.~\ref{fig:cross-main-before-after}a), only a subset of nodes are labeled as legitimate or phishing. Many neighbours of the central hub remain unlabeled (grey). Benign nodes (blue) form small, tightly connected groups. Malicious nodes (red) appear as thin branches at the periphery. Because the unlabeled nodes sit between these groups, it is difficult to understand the local risk pattern by inspection alone.

After applying SiNMULI  (Fig.~\ref{fig:cross-main-before-after}b), the previously unlabeled nodes are classified using edge signs inferred from directed triads. We apply our proposed algorithm~\ref{algo:sinumuli} to segregate the unlabeled URLs into malicious and benign. The resulting subgraph is more structured: benign and malicious pockets become clearly separated around the infrastructure hub. This view helps analysts see which regions of the local neighbourhood are safe, which clusters concentrate malicious activity, and how risk can propagate through shared infrastructure.

\subsection{Distribution of Triads}
In this section, we characterize the local signed structure of the SiNMULI network via \emph{triads}, i.e., the smallest motifs that encode balance/imbalance and status asymmetries in signed (and directed) graphs. Triads are theoretically well-motivated: classical structural balance theory predicts that certain sign patterns are more prevalent and stable than others, while modern treatments for directed settings emphasize status/ordering effects. Both of the prior frameworks use triads as their basic diagnostic unit. We quantify how often each label combination in the triads\{B = legitimate, M = malicious, U = unknown\}occurs. To understand skew better, we plot their frequencies on both linear and logarithmic scales. This triad census provides a local view of the network and complements the overall graph and community-level statistics by showing where imbalance or uncertainty tends to cluster. 
\par{\bf \em Triad filtering and counting protocol.}
We enumerate triads over the induced interaction graph and retain the seven informative node-label compositions that contain \emph{at least two labeled vertices}: 
I $(\mathrm{B},\mathrm{B},\mathrm{U})$, 
II $(\mathrm{B},\mathrm{B},\mathrm{M})$, 
III $(\mathrm{B},\mathrm{M},\mathrm{U})$, 
IV $(\mathrm{B},\mathrm{M},\mathrm{M})$, 
V $(\mathrm{M},\mathrm{M},\mathrm{U})$, 
VI $(\mathrm{B},\mathrm{B},\mathrm{B})$, 
VII $(\mathrm{M},\mathrm{M},\mathrm{M})$. 
We \emph{filter out} triads with insufficient label signal, i.e., $(\mathrm{B},\mathrm{U},\mathrm{U})$, $(\mathrm{M},\mathrm{U},\mathrm{U})$, and $(\mathrm{U},\mathrm{U},\mathrm{U})$, because balance/status predicates become indeterminate when fewer than two vertices are known. We also exclude degenerate cases (self-loops, missing edges) and collapse isomorphic label permutations via canonicalization, such that each composition is counted once. This protocol aligns with signed-network theory, where the balance test is defined over \emph{known} signed relations, and unknowns are handled via inference rather than included as evidence.

\par
\begin{figure}[t]
  \centering
  \includegraphics[width=6in, height=3in]{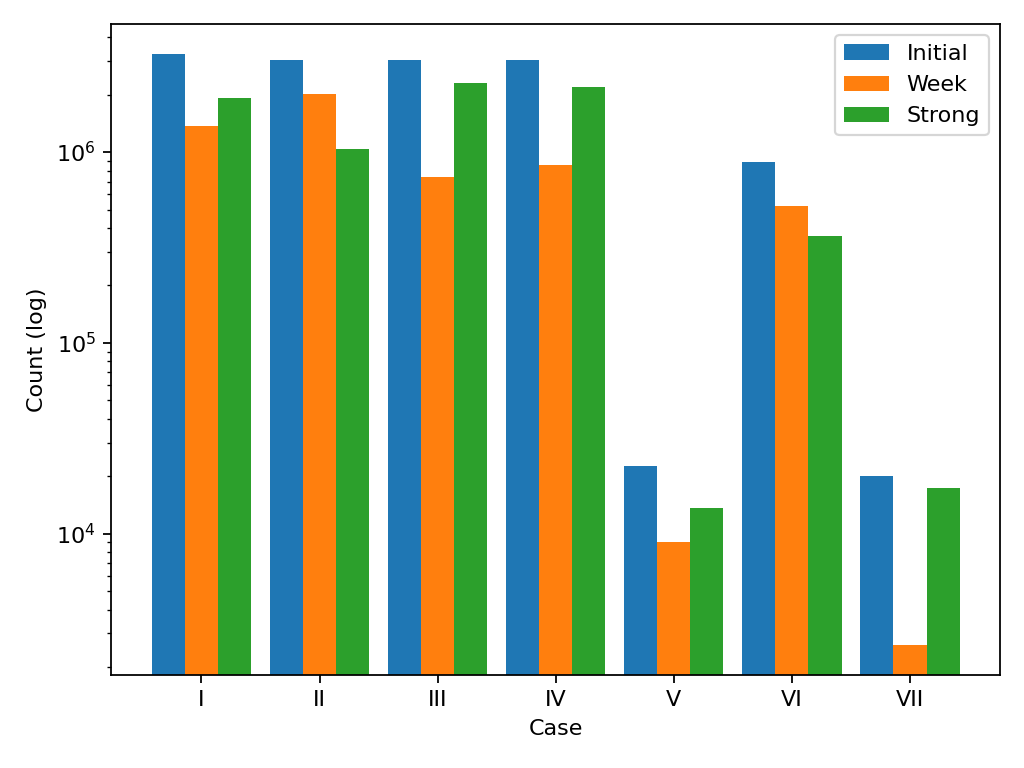}
 \caption{Figure shows the distribution of the seven distinct types of triads in the network. Along the x-axis, we plot each category, and along the y-axis, we plot the counts on a logarithmic scale.}
 \label{fig:triad-counts-log}
\end{figure}

 As shown in Fig.~\ref{fig:triad-counts-log}, mixed-type triads dominate the S{\em i}NMULI signed network. The \((B,B,M)\) pattern is most common (\(\approx 3.0\times 10^{13}\)), followed by \((B,M,M)\) (\(\approx 2.0\times 10^{13}\)); together they account for roughly 60--65\% of all triads. Purely benign \((B,B,B)\) triads occur at about \(1.5\times 10^{13}\), while purely malicious \((M,M,M)\) are rarer (\(\sim 0.8\times 10^{13}\)), indicating strongly balanced but less prevalent clusters. Triads involving unknown nodes---e.g., \((B,B,U)\) and \((M,M,U)\)---fall in the \(10^{12}\)--\(10^{13}\) range and contribute only marginally to overall balance. In short, benign--malicious interactions shape the network's typical motif, producing semi-stable, tension-prone structures that capture the system's structural imbalance.

\par On a logarithmic scale, triad counts span nearly three orders of magnitude, consistent with a heavy-tailed (scale-free) structure. The sharp drop from \((B,B,M)\) to \((B,B,B)\) shows that benign--malicious interlinks are far more common than purely cooperative clusters---an observation with direct implications for anomaly detection and propagation. The prevalence of \((B,M,M)\) and \((B,B,M)\) suggests a tendency toward weakly balanced states, where local inconsistencies coexist with global stability; these triads often act as ``tension hubs'' where polarity flips or edge re-evaluation occur during inference. By contrast, the relative scarcity of \((B,B,B)\) and \((M,M,M)\) indicates that strongly balanced communities appear as isolated cores rather than dominant formations. Quantitatively, these patterns support the \textsc{SiNMULI} hypothesis: real-world malicious ecosystems stabilise in partially balanced configurations, so weak triadic balance governs macro-dynamics and underpins triad-driven inference and sign propagation.

We further examine these seven triad types through the lens of strong and weak balance. By construction, triads of type (B,B,B) are strongly balanced under the Cartwright--Harary definition, since all three edges are positive. In contrast, (M,M,M) triads correspond to the all-negative configuration, which is unbalanced under strong balance but permitted under Davis's weak balance and therefore models hostile but internally stable clusters. Mixed triads such as (B,B,M) and (B,M,M) admit both strongly and weakly balanced edge-sign patterns: a single negative edge attached to the malicious node yields a strongly balanced (+,+,\ensuremath{-}) configuration, whereas sign assignments with three negative edges are only weakly balanced. In our inference pipeline, SiNMULI tends to favour sign assignments that move frequently occurring mixed triads toward these balanced configurations, either strong or weak. This behaviour is consistent with structural balance theory and provides a principled explanation for why triad-driven sign propagation stabilises quickly in the signed hyperlink network.

The triad-level analysis reveals a clear separation between structurally weak and structurally strong configurations. Triads of type II and VI predominantly instantiate weakly balanced patterns, in which the combination of edge signs fails to stabilise a consistent trust or distrust assignment across the three vertices and remains sensitive to small perturbations. In contrast, triads of type I, III, IV, V, and VII systematically generate strongly balanced cases, where the induced sign structure is internally consistent and reinforces either cohesive benign clusters or cohesive malicious clusters. This distribution implies that only a narrow subset of triad types acts as a source of local instability, while the majority of observed motifs contribute to the emergence of robustly polarized communities that drive reliable label propagation over the signed network.

\subsection{Evaluation Metrics}In our work, we have used several metrics to measure the performance of our proposed model as follows:
\paragraph{\bf{\emph{Confusion Matrix}}}The assessment of the proposed signed hyperlink classifier's effectiveness in predicting the malicious URLs from the collected dataset containing unlabeled URLs can be conducted through the utilization of a confusion matrix as depicted in the following: 
\begin{center}
\begin{tabular}{ |c|c|c| } 
 \hline
  & Malicious & Benign \\
 \hline
 Malicious & TP & FP \\ 
 \hline
 Benign & FN & TN \\ 
 \hline
\end{tabular}
\end{center}
Where, 
\begin{enumerate}
   \item \textbf{TP}= Total number of Malicious URLs correctly classified as Malicious.
   \item \textbf{TN}=Total number of Benign URLs correctly classified as Benign.
   \item \textbf{FP}=Total number of Benign URLs misclassified as Malicious.
   \item \textbf{FN}=Total number of Malicious URLs misclassified as Benign.
 \end{enumerate}
\paragraph{Accuracy} Accuracy measures the percentage of correct predictions the algorithm makes on a given dataset. Mathematically, accuracy is calculated by dividing the number of correct predictions by the total number of predictions:
\begin{equation} \label{eqn_Accu}
	Accuracy = \frac{TN + TP}{TP+FN+FP+TN} 
\end{equation}
\paragraph{Precision}Precision is a performance metric that evaluates the performance of a classification algorithm, specifically for positive class predictions. Mathematically, precision is calculated as:
\begin{equation} \label{eqn_prec}
	Precision = \frac{TP}{TP+FP} 
\end{equation}
\paragraph{Recall}Recall is a performance metric that measures the ability of the model to correctly identify positive instances or the proportion of actual positive instances that are correctly identified by the algorithm. The formula to calculate recall is as follows:
\begin{equation} \label{eqn_recall}
	Recall = \frac{TP}{TP+FN} 
\end{equation}
\paragraph{F1-score} F1-score combines precision and recall into a single value to provide a balanced assessment of the algorithm's effectiveness. The F1-score is the harmonic mean of precision and recall, calculated using the following formula:
\begin{equation} \label{eqn_f1}
	F_1 = \frac{2\times (Precision\times Recall)}{Precision + Recall} 
\end{equation}
\paragraph{ROC}The Receiver Operating Characteristic (ROC) curve is a graphical representation of the performance of a binary classification algorithm. 
\begin{equation} \label{eqn_TPR}
	TPR = \frac{TP}{TP+FN} 
\end{equation}
\begin{equation} \label{eqn_FPR}
	FPR = \frac{TN}{TN+FP} 
\end{equation}
\paragraph{AUC}In machine learning, the AUC (Area Under the Curve) is a metric used to evaluate the performance of binary classification models, typically in the context of a Receiver Operating Characteristic (ROC) curve. We summarize the performance of our proposed model in table~\ref{tab:perf_confusion}. In this table, we show the confusion matrix, which mentions the number of malicious and benign URLs that were tested to measure the performance. The results also show the performance of the model based on the standard performance metric.


\par The proposed signed network approach offers several notable advantages over traditional machine learning and deep learning-based systems:--{\bf \em a) Interpretability:} The model leverages sociological balance theory and hyperlink structure, which makes the inference process transparent---an essential feature for cybersecurity practitioners. {\bf \em b) Low Resource Requirements:}Unlike models such as DeepBF~\cite{patgiri2021}, or LBP-based models~\cite{guo2025}, which require labeled training data and computational overhead, our method operates with minimal training and relies on existing network structures. {\bf \em c) Robustness to Obfuscation:}Content- or URL-string-based approaches (e.g., \cite{le2018}, \cite{ma2009}) are often fooled by adversarial patterns. Our signed network-based approach is more resilient since hyperlink behaviour is harder to spoof at scale. {\bf \em d) Effective for Unknown Nodes:}Through majority-rule and triad reasoning, our system can label previously unseen domains without needing re-training or additional labeled samples. Table~\ref{tab:comparison_table} summarizes the key characteristics of SiNMULI relative to existing approaches across multiple evaluation dimensions.

\begin{table*}[!ht]
\centering
\scriptsize
\caption{Comparison of SiNMULI with Existing Malicious URL Detection Approaches}
\label{tab:comparison_table}
\begin{tabular}{|l|l|p{.4in}|c|c|c|c|}
\hline
\textbf{Approach} & \textbf{Feature Dependency} & \textbf{Inter-portability} & \makecell{\textbf{Adversarial}\\ \textbf{Robustness}} & \textbf{Scalability} & \makecell{\textbf{New Domain}\\ \textbf{Handling}} & \textbf{Complexity} \\
\hline
\textbf{SiNMULI (Proposed)} & Structural (Signed Graph) & High & High & High & Strong (51\% rule) & Low (Rule-based, no training) \\
DeepBF~\cite{patgiri2021} & Learned Bloom + CNN & Medium & Medium & Medium & Medium & Medium (Hybrid filter + CNN) \\
URLNet~\cite{le2018} & Lexical (Deep CNN) & Low & Low & Medium & Medium & High (Deep NLP pipeline) \\
LBP-GNN~\cite{guo2025} & Graph + IP/Host-level & Low & High & Medium & Strong & High (GNN + Message Passing) \\
Traditional ML~\cite{joshi2019} & Lexical \& Host-based & Medium & Low & High & Weak & Low (Shallow models + features) \\
Blacklist Methods~\cite{ma2009} & Manual Reports & High & Low & High & Very Weak & Very Low (Lookup only) \\
\hline
\end{tabular}
\end{table*}

\subsection{Comparison with Existing Models}

To fully place SiNMULI in the changing world of detecting bad URLs, we compare it with other methods that are already in use, including traditional machine learning, deep learning, and graph-based approaches. To perform this comparative study, we selected $6$ different existing modes: 1) Black-White List method~\cite{ma2009} 2) Traditional ML Model~\cite{joshi2019} 3) DeeBF~\cite{patgiri2021} 4) URLNet~\cite{le2018} 5) LBP-GNN~\cite{guo2025}. 


\begin{table*}[t]
\centering
\caption{Performance comparison of SiNMULI with baseline models and confusion matrix of predictions.}
\label{tab:perf_confusion}

\begin{minipage}[t]{0.62\textwidth}
\centering
\footnotesize     
\setlength{\tabcolsep}{5.2pt}
\renewcommand{\arraystretch}{1.4}

\textbf{(a) Performance comparison of SiNMULI with baseline models}

\begin{tabularx}{\linewidth}{l r r r r r}
\toprule
\textbf{Approach} &
\textbf{Acc (\%)} &
\textbf{Prec (\%)} &
\textbf{Rec (\%)} &
\textbf{F1 (\%)} &
\textbf{AUC} \\
\midrule
SiNMULI (Proposed) & 99.89 & 99.99 & 99.62 & 99.80 & 0.9981 \\
URLNet             & 97.63 & 91.87 & 96.02 & 93.90 & 0.9914 \\
DeepBF             & 97.98 & 91.98 & 97.90 & 94.85 & 0.9936 \\
Traditional ML     & 59.79 & 30.12 & 84.27 & 44.38 & 0.6471 \\
LBP-GNN            & 58.41 & 19.65 & 38.36 & 25.99 & 0.6412 \\
Blacklist          & 85.91 & 100.00 & 26.00 & 41.26 & 0.6300 \\
\bottomrule
\end{tabularx}
\end{minipage}
\hfill
\begin{minipage}[t]{0.34\textwidth}
\centering
\footnotesize
\renewcommand{\arraystretch}{1.3}

\textbf{(b) Confusion Matrix of SiNMULI}

\vspace{5pt}

\begin{tabular}{c|c|c}
\toprule
 & \textbf{Mal.} & \textbf{Ben.} \\
\midrule
\textbf{Mal.} & 12958 & 50 \\
\textbf{Ben.} & 1 & 33175 \\
\bottomrule
\end{tabular}

\end{minipage}

\end{table*}










Table~\ref{tab:comparison_table} summarizes the end-to-end performance of SiNMULI relative to six representative baselines spanning blacklist lookup, traditional lexical/host-feature models, deep lexical CNNs, and graph neural approaches. Across all metrics, SiNMULI achieves near-perfect results (Precision = 99.99\%, Recall = 99.62\%, F1 = 99.80\%, AUC = 0.9981), while maintaining a competitive computational cost per URL. In particular, SiNMULI consistently matches or exceeds the AUC of deep lexical models (URLNet, DeepBF) and graph-based baselines (LBP-GNN), despite relying on lightweight signed-network inference rather than heavy end-to-end training.


\begin{figure*}[t]
\centering

\begin{minipage}[t]{0.48\textwidth}
\centering
\includegraphics[width=\linewidth, height=2.3in]{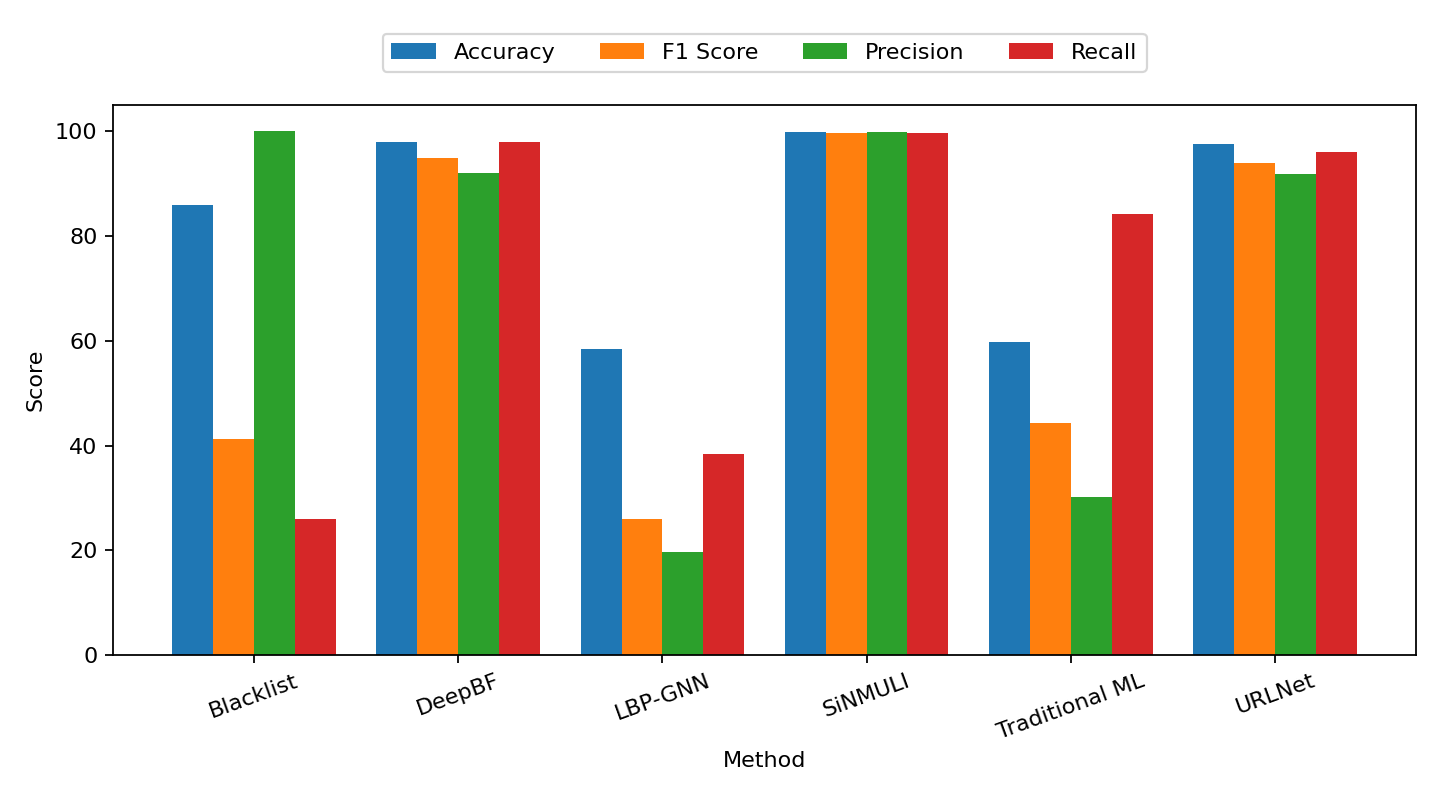}
\caption*{\textbf{(a)} Accuracy, F1-score, Precision and Recall comparison of S{\em i}NMULI with baseline models.}
\label{fig:heatmap_comparison}
\end{minipage}
\hfill
\begin{minipage}[t]{0.48\textwidth}
\centering
\includegraphics[width=\linewidth]{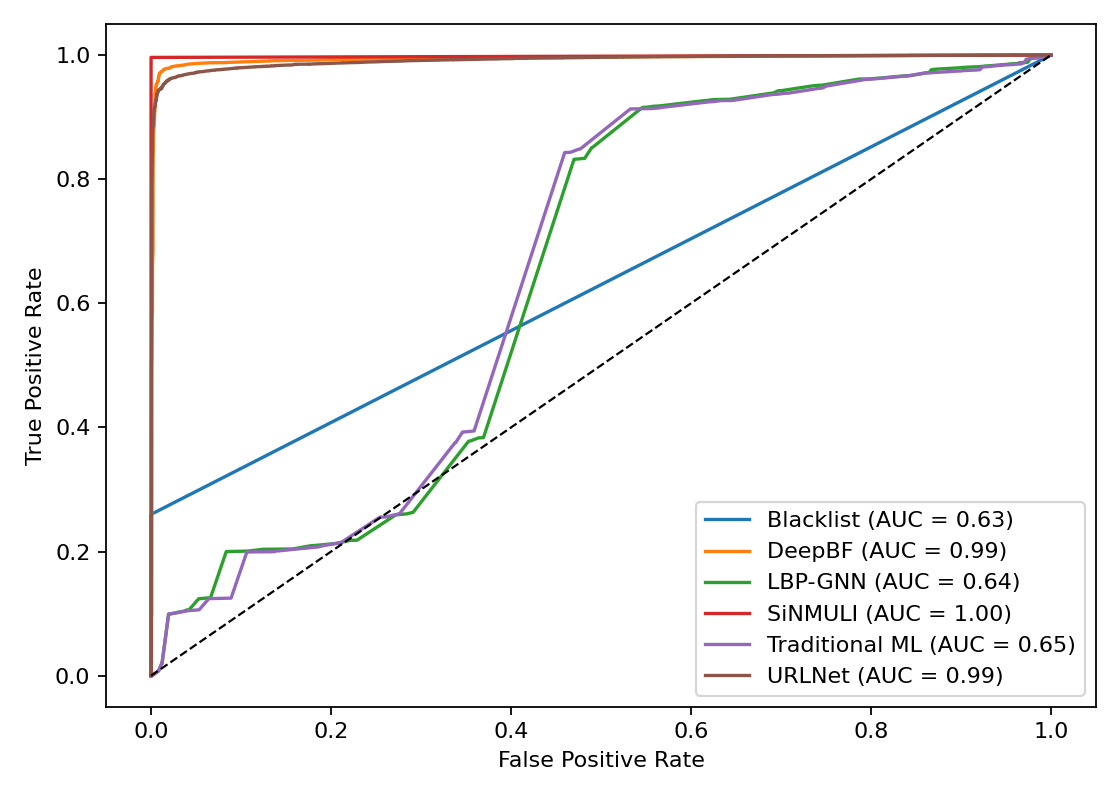}
\caption*{\textbf{(b)} ROC comparison of S{\em i}NMULI with baseline models.}
\label{fig:ROC}
\end{minipage}

\caption{Performance visualization of the proposed S{\em i}NMULI model:  
(a) Accuracy, Precision, Recall and F1-score comparison across models;  
(b) ROC curve comparison with baseline approaches.}
\label{fig:model_performance}
\end{figure*}


\par{\bf \em Insight of the Performace Comparison: } SiNMULI delivers strong, near-perfect performance across our experiments: \textbf{Accuracy} $=0.9989$, \textbf{Precision} $=0.9999$, \textbf{Recall} $=0.9962$, and \textbf{F1} $=0.9980$. The confusion matrix (\(\mathrm{TP}=12{,}958\), \(\mathrm{TN}=33{,}175\), \(\mathrm{FP}=1\), \(\mathrm{FN}=50\)) implies a negligible false-positive rate (\(\approx 3\times10^{-5}\)) alongside high sensitivity, yielding a balanced error profile suitable for deployment. The ROC curve in Fig.~\ref{fig:model_performance}(b) places SiNMULI on the upper performance envelope relative to baselines across operating points, showing consistently higher true-positive rates at comparable false-positive rates. We attribute these gains to the explicit use of \emph{signed} hyperlink structure and balance-theoretic triadic reasoning, which propagate trust/distrust signals to confidently label weakly connected or previously unseen nodes, and to the framework's low training overhead that leans on network structure rather than large labeled corpora---traits that also improve robustness to content obfuscation (as shown in Fig.~\ref{fig:model_performance}(a) and Table~\ref{tab:perf_confusion}). Collectively, the results indicate that structure-aware, multi-layer signed modeling yields reliable, generalisable, and resource-efficient malicious-URL detection that outperforms the traditional ML, deep lexical, and graph baselines reported alongside SiNMULI.

\subsection{Application in Industrial and Enterprise Environments}
Industrial control systems and large enterprise networks increasingly rely
on web-based update channels (e.g., firmware repositories, vendor portals,
SCADA configuration endpoints. In such environments, misclassified or
Compromised URLs can directly translate into plant downtime or safety issues
incidents. SiNMULI's lightweight, training-free design makes it suitable
for deployment on industrial gateways and secure web proxies, where it can
continuously monitor hyperlink interactions among update servers, mirrors,
and internal hosts.

Because SiNMULI operates on local triads and majority rules over edges, much of the inference can be pushed to the edge of the network. Routers, industrial gateways, or SDN switches can maintain local signed subgraphs and apply balance-based updates incrementally as new URLs and links appear, forwarding only aggregated trust scores to a central controller. This distributed design reduces bandwidth overhead and enables low-latency decisions close to where traffic is observed.

For example, URLs involved in firmware distribution often form a tightly
knit a benign cluster in the signed hyperlink graph; sudden emergence of
links to low-trust domains or previously unseen external hosts can be
flagged and scored by SiNMULI, yielding an interpretable risk signal for
operators before an update is rolled out fleet-wide.

\section{Analysis of Malicious Sites}\label{s:mal_analysis} 

We analyze malicious sites in detail because a realistic detector cannot be designed in the abstract: it must be grounded in how phishing and other attacks are actually realized in the wild. Systematic characterization of our malicious HTML corpus along lexical, content/DOM, and hyperlink-structure axes exposes the concrete design choices of attackers, such as how they shape URLs, structure forms and scripts, and wire pages into the web graph. This, in turn, tells us where the most discriminative and stable signals live, which features from prior work remain useful, and which additional cues are needed to handle modern kits, obfuscation, and template reuse. By quantifying these patterns rather than assuming them, the analysis reduces feature engineering to an evidence-driven process, helps avoid overfitting to URL-only heuristics or brittle blacklists, and ultimately supports the construction of hybrid models that generalize better to previously unseen lures and evolving malicious infrastructures. We have summarized all the features that are to be analyzed from the malicious URL perspective (listed in table~\ref{tab:mal_features}).

This section characterizes our HTML corpus along three complementary axes that are widely used in phishing analysis: (i) \emph{lexical} signals derived from the page URL (length/entropy, special characters, subdomain depth, tokens), (ii) \emph{content/DOM} cues (document size, forms and input controls, scripting/iframes, inline/base64 assets), and (iii) \emph{hyperlink structure} (internal vs.\ external links, unique outbound domains, empty and \texttt{mailto:} anchors). Jointly, these views capture both the page's surface-level appearance and the ways it connects to the web graph---two facets that prior work has repeatedly shown to be complementary. Content-based detectors such as \mbox{CANTINA} motivated DOM/text features, while later hybrid systems explicitly fuse URL-, HTML/JS-, and link-based indicators to improve generalization beyond blacklist matching. We adopt the same rationale here and report descriptive statistics to guide feature selection for downstream models.

\begin{figure}[hbt]
  \centering
  \includegraphics[width=\linewidth, height=0.2\textheight]{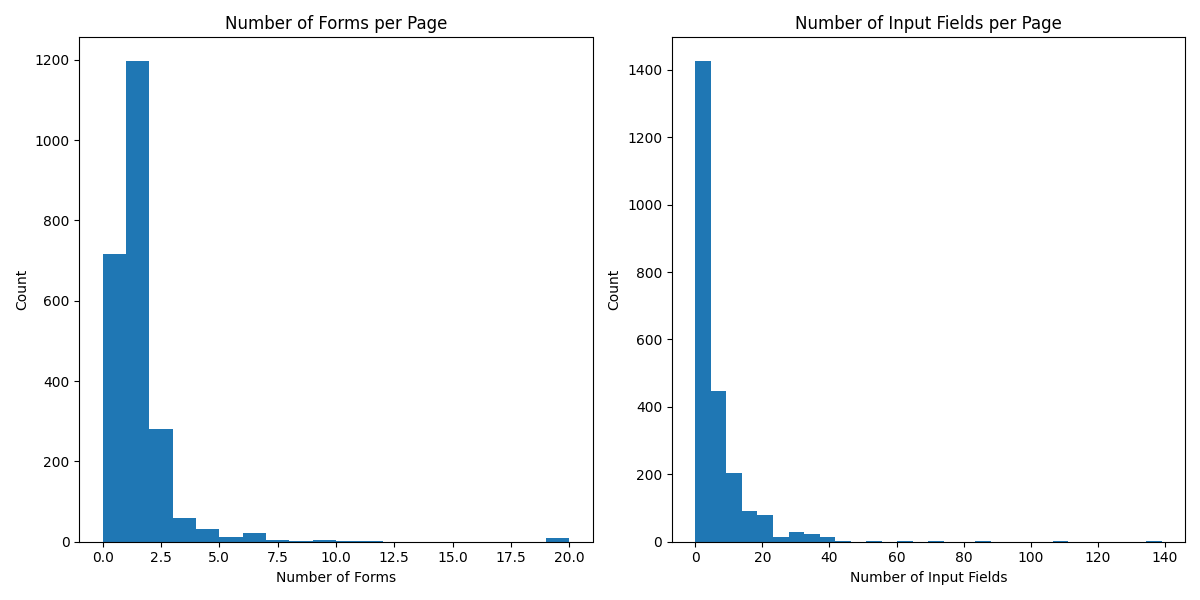}
  \vspace{-.2in}
  \caption{Left: histogram of forms per page. Right: histogram of input fields per page. Both are highly right-skewed.}
  \label{fig:forms}
\end{figure}

\begin{figure*}[hbt]
    \centering
    
    \begin{subfigure}{0.32\textwidth}
        \centering
        \includegraphics[width=\linewidth]{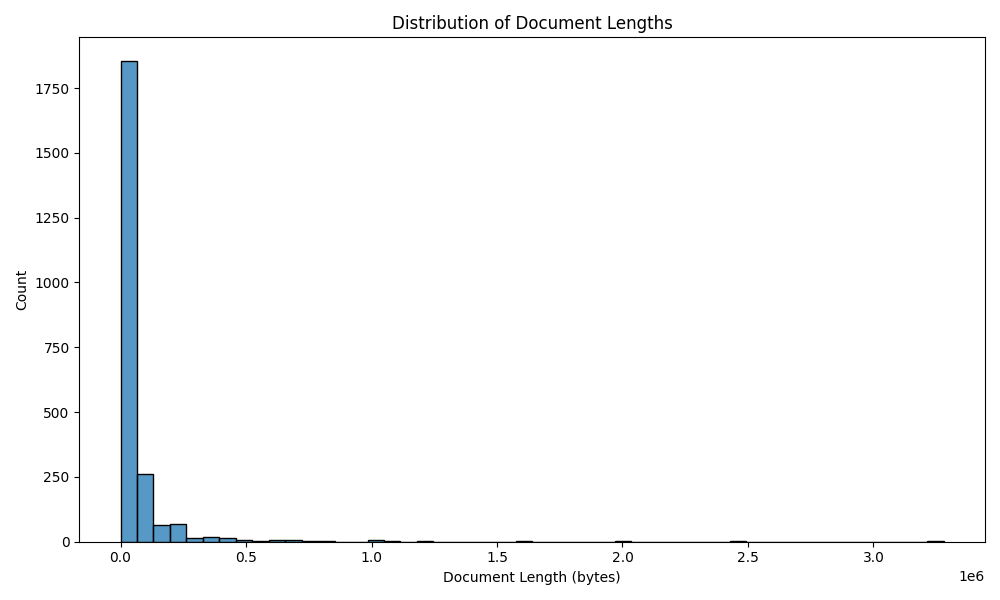}
        \caption{Distribution of doc. lengths (bytes).}
        \label{fig:doclen}
    \end{subfigure}
    \hfill
    \begin{subfigure}{0.32\textwidth}
        \centering
        \includegraphics[width=\linewidth]{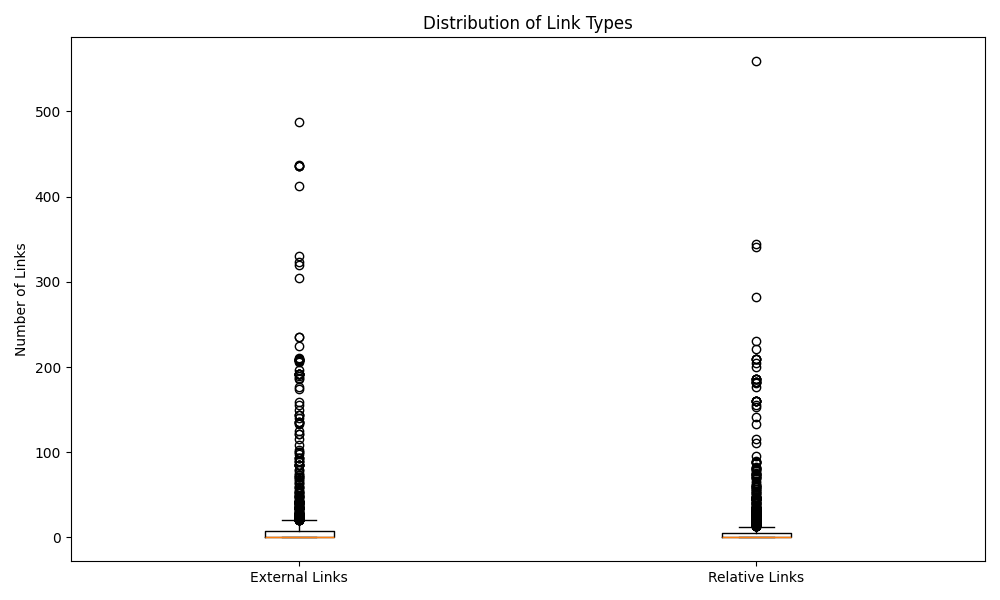}
        \caption{External vs.\ relative link distri.}
        \label{fig:links}
    \end{subfigure}
    \hfill
    \begin{subfigure}{0.32\textwidth}
        \centering
        \includegraphics[width=\linewidth]{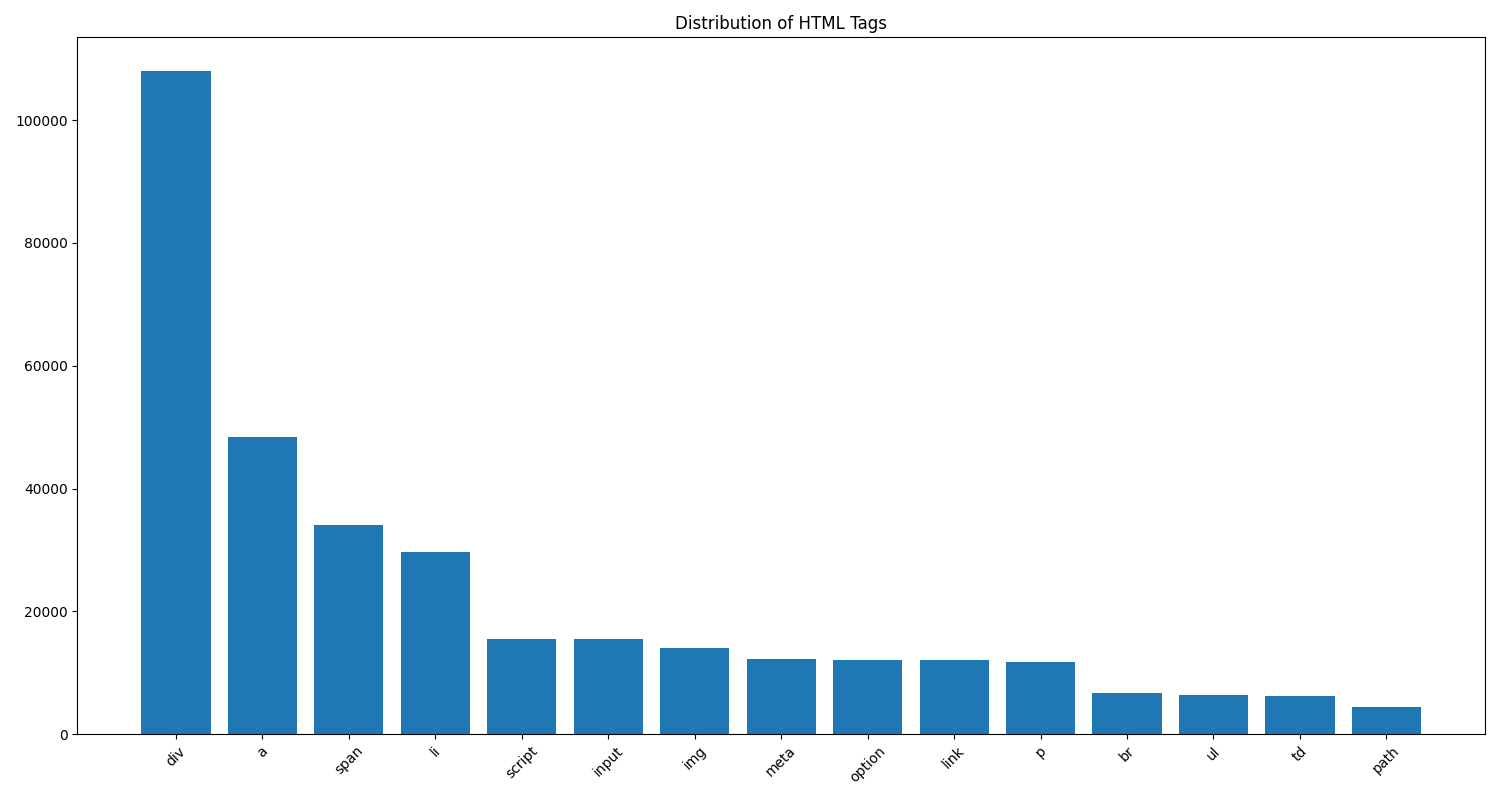}
        \caption{Top HTML tags distribution.}
        \label{fig:tags}
    \end{subfigure}

    \caption{Figure~\ref{fig:doclen},  Distribution of document lengths (bytes) across the corpus. Along the x-axis, we plot document length measured in bytes, whereas the y-axis represents the raw count. In figure~\ref{fig:links}, we show external vs.\ relative link distributions. Figure~\ref{fig:tags} shows the distribution of top HTML tags across pages. Along the x-axis, we plot the top tags, and the  y-axis plots the raw count of that corresponding tag.}
    \label{fig:three_row}
\end{figure*}

Document size is highly right-skewed (Fig.~\ref{fig:doclen}): most pages are small, with a long tail of large files---often templates with inlined assets or obfuscated bundles. Forms and input controls exhibit a sparse core (typically $0$--$2$ per page) and a thin tail of complex pages (Fig.~\ref{fig:forms}); for phishing, the \emph{type} and \emph{destination} of forms matter more than raw counts (e.g., password fields and off-domain handlers). Hyperlink boxplots reveal many pages with few links and a minority with hundreds (Fig.~\ref{fig:links}); separating internal from external links is useful because pages that push users off-domain (and whose forms post off-domain) are riskier in aggregate. Finally, the tag mix (Fig.~\ref{fig:tags}) is dominated by layout/navigation elements (\texttt{div}, \texttt{a}, \texttt{span}); variance in \texttt{script}, \texttt{input}, \texttt{iframe}, and \texttt{link}/\texttt{meta} is especially informative because that is where credential capture, obfuscation, and third-party dependencies typically live. Beyond detection, the external-link heavy tail is operationally relevant: large-scale studies show nontrivial \emph{link rot} over time, implying that pages with heavy external dependencies are more brittle to reproduce.
\vspace{-.1in}

\noindent\textbf{Implications for feature design and detection.}
These distributions support a hybrid feature set: lexical URL cues (e.g., host as IP, ``@'', redirection ``//'', hyphens, token matches like \emph{login}/\emph{verify}, shorteners), content/DOM heuristics (right-click disabling, pop-ups, \texttt{onmouseover}, base64 density, number and type of forms), and hyperlink/SFH signals (internal vs.\ external link ratio, unique outbound domains, \emph{server form handler} domain compared to page domain, \texttt{mailto:} submissions). Public benchmarks (e.g., the UCI ``Phishing Websites''/``Website Phishing'' datasets) encode many of these exact attributes---\texttt{SFH}, \texttt{Request\_URL}, \texttt{URL\_of\_Anchor}, \texttt{Links\_in\_tags}, \texttt{popUpWindow}, \texttt{having\_IP\_Address}---and surveys consistently report that models combining URL + HTML/JS + link features outperform URL-only baselines, particularly on look-alike domains and previously unseen lures. In our corpus, we therefore prioritize \emph{where} forms submit, the external-resource/link profile, and the presence of obfuscating script patterns alongside standard URL lexicals; together these yield transparent, engineerable signals that align with prior art and explain the outliers seen in Figures~\ref{fig:forms}--~\ref{fig:three_row}.

\begin{table*}[h]
\small
\centering
\caption{Summary of analysis features for malicious sites and their relevance to detection.}
\label{tab:mal_features}
\begin{tabularx}{\textwidth}{p{0.17\textwidth} p{0.33\textwidth} X}
\toprule
\textbf{Axis / feature group} & \textbf{Representative signals} & \textbf{Why it matters for detection} \\
\midrule
Lexical URL structure &
URL length, character entropy, subdomain depth, use of separators and repeated ``//''. &
Phishing URLs often use long, noisy, deeply nested paths to hide impersonation and tracking parameters, diverging from typical benign URL structure. \\[0.25em]

Host representation \& redirection tricks &
Host as raw IP, URL shorteners, unusual TLDs, ``@'' and pseudo-redirect segments. &
Raw IP hosts, shorteners, and misleading separators are common in transient malicious infrastructure and help expose the true origin of the page. \\[0.25em]

Semantic URL tokens &
Presence of tokens such as \emph{login}, \emph{verify}, \emph{update}, \emph{secure}, brand names. &
Security- and brand-related terms are over-represented in credential-theft and verification lures and become more discriminative when combined with structural URL cues. \\[0.25em]

Document size \& inlined assets &
HTML size, number/size of inline images and scripts, base64-encoded content. &
Very large pages or high base64 density often indicate obfuscated bundles or reusable phishing templates rather than ordinary, modular benign pages. \\[0.25em]

Forms and input controls &
Counts of forms, presence of password fields, server form handler (SFH) domain vs.\ page domain. &
Credential harvesting requires specific inputs and off-domain handlers; mismatches between page and SFH domains (e.g., \texttt{SFH}) are strong phishing indicators. \\[0.25em]

Scripting and UI behaviour &
Number of \texttt{script} tags, pop-up windows, right-click disabling, event handlers such as \texttt{onmouseover}. &
Aggressive or anti-inspection UI patterns and obfuscated scripts are more prevalent on malicious pages and capture behaviours not visible from the URL alone. \\[0.25em]

Iframes and embedded widgets &
Use of \texttt{iframe} and similar embedding constructs, embedded login panels. &
Embedded widgets enable visual cloning of legitimate sites while redirecting credentials to attacker-controlled endpoints, providing a structural cloning signal. \\[0.25em]

Hyperlink structure (internal vs.\ external) &
Counts and ratios of on-domain vs.\ off-domain links, number of unique outbound domains. &
Atypical internal/external link ratios and many distinct outbound domains are characteristic of spam and phishing landing pages; related to features such as \texttt{URL\_of\_Anchor} and \texttt{Links\_in\_tags}. \\[0.25em]

Anchor semantics \& mailto targets &
Empty anchors, anchors pointing to \texttt{mailto:}, misleading or non-navigational links. &
Such anchors often appear in low-effort spam and contact-harvesting pages and help distinguish them from well-structured, benign navigation menus. \\[0.25em]

External resource profile \& brittleness &
Fraction of scripts/images/styles loaded from external domains; long external-link tail; reliance on a few critical endpoints. &
High external dependence (e.g., \texttt{Request\_URL}) and brittle link profiles are typical of short-lived malicious infrastructures and support longitudinal fingerprinting beyond simple blacklists. \\
\bottomrule
\end{tabularx}
\end{table*}




\section{Conclusions and Future Works}\label{s:conclusion}

In this paper, we presented SiNMULI, a signed-network--based framework for detecting malicious URLs. Starting from a labeled phishing dataset, we constructed a large-scale signed hyperlink graph where websites act as nodes and hyperlinks as directed trust/distrust edges. Using a directed balance theory tailored to adversarial web graphs, we inferred missing edge signs and classified previously unseen domains through a simple 51\% majority rule. Experimental evaluation shows that SiNMULI achieves near-perfect performance (accuracy 99.89\%, precision 99.99\%, Recall 99.62\%, F$_1$ 99.80\%), while remaining interpretable, training-free, and computationally efficient compared with machine-learning baselines. Our proposed model not only explores the social aspect of hyperlink generation but also explores the new horizon of malicious URL detection. Performance of S{\em i}MULI is compared with existing baselines, and it reveals that our proposed model performs better than existing ones. Our results also demonstrate that structural properties, e.g., lower centrality, sparse internal links, and weaker clustering, consistently differentiate phishing sites from legitimate ones. These findings confirm that signed connectivity patterns carry strong maliciousness signals and effectively answer our three research questions on model utility, efficiency, and balance-theoretic inference.

Despite its strong performance, S{\em i}NMULI has several limitations. Comprehensive data collection was difficult because many phishing websites are short-lived, block crawlers, or reveal only partial content. It results in incomplete link structures and isolated nodes in the graph. This incomplete coverage may obscure clusters or relationships within the malicious ecosystem. Thus, it restricts how well the model generalizes to live traffic. Further, modern websites often load essential content through scripts or user interactions that basic crawlers do not capture. Thus, it leads to additional missing links. Finally, the current pipeline operates offline: constructing the full signed network and running iterative balance-based label propagation are computationally costly, making real-time deployment challenging without substantial engineering improvements.

Future work can strengthen S{\em i}NMULI in several ways. First, developing an incremental or streaming version of the signed-network model would allow the system to update trust relationships dynamically as new pages and links appear. Thus, it would be able to improve the adaptability of the model to fast-moving phishing campaigns. Second, integrating network-structure signals with content-based features, e.g., text, images, or URL components, could enhance robustness in cases where link structure is sparse or ambiguous. Improving the efficiency of the model is also essential to cope with real-world scenarios. Thus, incorporating faster data structures, improved pruning strategies, and targeted crawling could reduce computation and enable near--real-time analysis on large web graphs. 

\section*{Declarations}
\begin{itemize}
\item \textbf{Funding}:
No funds, grants, or other support were received to conduct this study.
\item \textbf{Conflict of interest/Competing interests}:  On behalf of all authors, the corresponding author states that there is no conflict of interest.
\item \textbf{Data availability}: Publicly available.
\item \textbf{Author contribution}: {\bf Avijit Gayen:} Conceptualization, Methodology, Experiment, Writing. 
{\bf Sayan Mondal:} Data collection, Methodology, Experiment, Writing.  
{\bf Angshuman Jana:} Supervision, Writing review \& editing.  
\end{itemize}

\bibliographystyle{ACM-Reference-Format}
\bibliography{Bibliography1,Bibliography2}


\begin{thebibliography}{33}


\ifx \showCODEN    \undefined \def \showCODEN     #1{\unskip}     \fi
\ifx \showISBNx    \undefined \def \showISBNx     #1{\unskip}     \fi
\ifx \showISBNxiii \undefined \def \showISBNxiii  #1{\unskip}     \fi
\ifx \showISSN     \undefined \def \showISSN      #1{\unskip}     \fi
\ifx \showLCCN     \undefined \def \showLCCN      #1{\unskip}     \fi
\ifx \shownote     \undefined \def \shownote      #1{#1}          \fi
\ifx \showarticletitle \undefined \def \showarticletitle #1{#1}   \fi
\ifx \showURL      \undefined \def \showURL       {\relax}        \fi
\providecommand\bibfield[2]{#2}
\providecommand\bibinfo[2]{#2}
\providecommand\natexlab[1]{#1}
\providecommand\showeprint[2][]{arXiv:#2}

\bibitem[Anderson(2010)]%
        {anderson2010security}
\bibfield{author}{\bibinfo{person}{Ross~J. Anderson}.}
  \bibinfo{year}{2010}\natexlab{}.
\newblock \bibinfo{booktitle}{\emph{Security Engineering: A Guide to Building
  Dependable Distributed Systems}}.
\newblock \bibinfo{publisher}{John Wiley \& Sons}.
\newblock


\bibitem[{Anti-Phishing Working Group}(2023)]%
        {apwg2023}
\bibfield{author}{\bibinfo{person}{{Anti-Phishing Working Group}}.}
  \bibinfo{year}{2023}\natexlab{}.
\newblock \bibinfo{title}{Phishing Activity Trends Report, 2nd Quarter 2023}.
\newblock
\urldef\tempurl%
\url{https://docs.apwg.org/reports/apwg_trends_report_q2_2023.pdf}
\showURL{%
\tempurl}
\newblock
\shownote{Accessed: 2023-11-16}.


\bibitem[Ariyadasa et~al\mbox{.}(2021)]%
        {phishing_dataset}
\bibfield{author}{\bibinfo{person}{Subhash Ariyadasa}, \bibinfo{person}{Shantha
  Fernando}, {and} \bibinfo{person}{Subha Fernando}.}
  \bibinfo{year}{2021}\natexlab{}.
\newblock \bibinfo{title}{Phishing Websites Dataset}.
\newblock \bibinfo{howpublished}{\url{https://doi.org/10.17632/n96ncsr5g4.1}}.
\newblock
\href{https://doi.org/10.17632/n96ncsr5g4.1}{doi:\nolinkurl{10.17632/n96ncsr5g4.1}}


\bibitem[Bell and Komisarczuk(2020)]%
        {bell2020analysis}
\bibfield{author}{\bibinfo{person}{Simon Bell} {and} \bibinfo{person}{Peter
  Komisarczuk}.} \bibinfo{year}{2020}\natexlab{}.
\newblock \showarticletitle{An Analysis of Phishing Blacklists: Google Safe
  Browsing, OpenPhish, and PhishTank}. In \bibinfo{booktitle}{\emph{Proceedings
  of the Australasian Computer Science Week Multiconference}}.
  \bibinfo{pages}{1--11}.
\newblock


\bibitem[Cartwright and Harary(1956)]%
        {cartwright1956structural}
\bibfield{author}{\bibinfo{person}{Dorwin Cartwright} {and}
  \bibinfo{person}{Frank Harary}.} \bibinfo{year}{1956}\natexlab{}.
\newblock \showarticletitle{Structural Balance: A Generalization of Heider's
  Theory}.
\newblock \bibinfo{journal}{\emph{Psychological Review}} \bibinfo{volume}{63},
  \bibinfo{number}{5} (\bibinfo{year}{1956}), \bibinfo{pages}{277--293}.
\newblock
\href{https://doi.org/10.1037/h0046049}{doi:\nolinkurl{10.1037/h0046049}}


\bibitem[Chiang et~al\mbox{.}(2014)]%
        {chiang2014prediction}
\bibfield{author}{\bibinfo{person}{Kai-Yang Chiang}, \bibinfo{person}{Nikhil
  Natarajan}, \bibinfo{person}{Nikolaj Tatti}, {and}
  \bibinfo{person}{Inderjit~S. Dhillon}.} \bibinfo{year}{2014}\natexlab{}.
\newblock \showarticletitle{Prediction and Clustering in Signed Networks: A
  Local to Global Perspective}.
\newblock \bibinfo{journal}{\emph{Journal of Machine Learning Research}}
  \bibinfo{volume}{15}, \bibinfo{number}{1} (\bibinfo{year}{2014}),
  \bibinfo{pages}{1177--1213}.
\newblock


\bibitem[Gao et~al\mbox{.}(2020)]%
        {gao2020}
\bibfield{author}{\bibinfo{person}{Jian Gao} {et~al\mbox{.}}}
  \bibinfo{year}{2020}\natexlab{}.
\newblock \showarticletitle{Phishing URL Detection based on CNN and GRU}.
\newblock \bibinfo{journal}{\emph{IEEE Access}}  \bibinfo{volume}{8}
  (\bibinfo{year}{2020}), \bibinfo{pages}{25681--25688}.
\newblock


\bibitem[Gayen et~al\mbox{.}(2024)]%
        {Gayen2024}
\bibfield{author}{\bibinfo{person}{Avijit Gayen}, \bibinfo{person}{Sukriti
  Santra}, \bibinfo{person}{Ayush Dey}, {and} \bibinfo{person}{Angshuman
  Jana}.} \bibinfo{year}{2024}\natexlab{}.
\newblock \showarticletitle{A Network Model to Study the Underlying
  Characteristics of Phishing Sites}.
\newblock \bibinfo{journal}{\emph{IEEE Networking Letters}}
  \bibinfo{volume}{6}, \bibinfo{number}{1} (\bibinfo{year}{2024}),
  \bibinfo{pages}{46--49}.
\newblock
\href{https://doi.org/10.1109/LNET.2023.3347724}{doi:\nolinkurl{10.1109/LNET.2023.3347724}}


\bibitem[Gibbs et~al\mbox{.}(2018)]%
        {gibbs2018introduction}
\bibfield{author}{\bibinfo{person}{Paul Gibbs}, \bibinfo{person}{Linda
  Neuhauser}, {and} \bibinfo{person}{Dena Fam}.}
  \bibinfo{year}{2018}\natexlab{}.
\newblock \showarticletitle{Introduction--The Art of Collaborative Research and
  Collective Learning: Transdisciplinary Theory, Practice and Education}.
\newblock In \bibinfo{booktitle}{\emph{Transdisciplinary Theory, Practice and
  Education}}. \bibinfo{publisher}{Springer}, \bibinfo{pages}{3--9}.
\newblock


\bibitem[Guo et~al\mbox{.}(2025)]%
        {guo2025}
\bibfield{author}{\bibinfo{person}{W. Guo} {et~al\mbox{.}}}
  \bibinfo{year}{2025}\natexlab{}.
\newblock \showarticletitle{Efficient Phishing URL Detection Using Graph-based
  Machine Learning and Loopy Belief Propagation}.
\newblock \bibinfo{journal}{\emph{arXiv preprint arXiv:2501.06912}}
  (\bibinfo{year}{2025}).
\newblock
\showeprint[arxiv]{2501.06912}~[cs.CR]


\bibitem[Heider(1946)]%
        {heider1946attitudes}
\bibfield{author}{\bibinfo{person}{Fritz Heider}.}
  \bibinfo{year}{1946}\natexlab{}.
\newblock \showarticletitle{Attitudes and Cognitive Organization}.
\newblock \bibinfo{journal}{\emph{Journal of Psychology}} \bibinfo{volume}{21},
  \bibinfo{number}{1} (\bibinfo{year}{1946}), \bibinfo{pages}{107--112}.
\newblock


\bibitem[Insko(1984)]%
        {INSKO198489}
\bibfield{author}{\bibinfo{person}{Chester~A. Insko}.}
  \bibinfo{year}{1984}\natexlab{}.
\newblock \showarticletitle{Balance Theory, The Jordan Paradigm, and The Wiest
  Tetrahedron}.
\newblock \bibinfo{series}{Advances in Experimental Social Psychology},
  Vol.~\bibinfo{volume}{18}. \bibinfo{publisher}{Academic Press},
  \bibinfo{pages}{89--140}.
\newblock
\showISSN{0065-2601}
\href{https://doi.org/10.1016/S0065-2601(08)60143-4}{doi:\nolinkurl{10.1016/S0065-2601(08)60143-4}}


\bibitem[Joshi et~al\mbox{.}(2019)]%
        {joshi2019}
\bibfield{author}{\bibinfo{person}{Anupam Joshi} {et~al\mbox{.}}}
  \bibinfo{year}{2019}\natexlab{}.
\newblock \showarticletitle{Using Lexical Features for Malicious URL Detection:
  A Machine Learning Approach}.
\newblock \bibinfo{journal}{\emph{arXiv preprint arXiv:1910.06277}}
  (\bibinfo{year}{2019}).
\newblock
\showeprint[arxiv]{1910.06277}~[cs.CR]


\bibitem[Khonji et~al\mbox{.}(2013)]%
        {Khonji2013Survey}
\bibfield{author}{\bibinfo{person}{M. Khonji}, \bibinfo{person}{Y. Iraqi},
  {and} \bibinfo{person}{A. Jones}.} \bibinfo{year}{2013}\natexlab{}.
\newblock \showarticletitle{Phishing Detection: A Literature Survey}.
\newblock \bibinfo{journal}{\emph{IEEE Communications Surveys \& Tutorials}}
  \bibinfo{volume}{15}, \bibinfo{number}{4} (\bibinfo{year}{2013}),
  \bibinfo{pages}{2091--2121}.
\newblock
\href{https://doi.org/10.1109/SURV.2013.032213.00009}{doi:\nolinkurl{10.1109/SURV.2013.032213.00009}}


\bibitem[Kim et~al\mbox{.}(2022)]%
        {kim2022}
\bibfield{author}{\bibinfo{person}{T. Kim} {et~al\mbox{.}}}
  \bibinfo{year}{2022}\natexlab{}.
\newblock \showarticletitle{Phishing URL Detection: A Network-based Approach
  Robust to Evasion}.
\newblock \bibinfo{journal}{\emph{IEEE Transactions on Information Forensics
  and Security}} (\bibinfo{year}{2022}).
\newblock


\bibitem[Kou et~al\mbox{.}(2020)]%
        {kou2020}
\bibfield{author}{\bibinfo{person}{H. Kou} {et~al\mbox{.}}}
  \bibinfo{year}{2020}\natexlab{}.
\newblock \showarticletitle{Trust-based Missing Link Prediction in Signed
  Social Networks}.
\newblock \bibinfo{journal}{\emph{ACM Transactions on Knowledge Discovery from
  Data}} \bibinfo{volume}{14}, \bibinfo{number}{3} (\bibinfo{year}{2020}),
  \bibinfo{pages}{1--27}.
\newblock


\bibitem[Kumar et~al\mbox{.}(2020)]%
        {kumar2020novel}
\bibfield{author}{\bibinfo{person}{Abhishek Kumar}, \bibinfo{person}{Jyotir~Moy
  Chatterjee}, {and} \bibinfo{person}{Vicente~Garc{\'\i}a D{\'\i}az}.}
  \bibinfo{year}{2020}\natexlab{}.
\newblock \showarticletitle{A novel hybrid approach of SVM combined with NLP
  and probabilistic neural network for email phishing}.
\newblock \bibinfo{journal}{\emph{International Journal of Electrical and
  Computer Engineering}} \bibinfo{volume}{10}, \bibinfo{number}{1}
  (\bibinfo{year}{2020}), \bibinfo{pages}{486}.
\newblock


\bibitem[Le et~al\mbox{.}(2018)]%
        {le2018}
\bibfield{author}{\bibinfo{person}{Hai Le}, \bibinfo{person}{Minh Pham},
  \bibinfo{person}{Doyen Sahoo}, {and} \bibinfo{person}{Steven~C. Hoi}.}
  \bibinfo{year}{2018}\natexlab{}.
\newblock \showarticletitle{URLNet: Learning a URL Representation with Deep
  Learning for Malicious URL Detection}. In
  \bibinfo{booktitle}{\emph{Proceedings of the 32nd AAAI Conference on
  Artificial Intelligence}}. \bibinfo{pages}{7950--7957}.
\newblock


\bibitem[Leskovec et~al\mbox{.}(2010)]%
        {leskovec2010signed}
\bibfield{author}{\bibinfo{person}{Jure Leskovec}, \bibinfo{person}{Daniel
  Huttenlocher}, {and} \bibinfo{person}{Jon Kleinberg}.}
  \bibinfo{year}{2010}\natexlab{}.
\newblock \showarticletitle{Signed Networks in Social Media}. In
  \bibinfo{booktitle}{\emph{Proceedings of the SIGCHI Conference on Human
  Factors in Computing Systems (CHI)}}. \bibinfo{publisher}{ACM},
  \bibinfo{pages}{1361--1370}.
\newblock
\href{https://doi.org/10.1145/1753326.1753532}{doi:\nolinkurl{10.1145/1753326.1753532}}


\bibitem[Ma et~al\mbox{.}(2009)]%
        {ma2009}
\bibfield{author}{\bibinfo{person}{Justin Ma}, \bibinfo{person}{Lawrence~K.
  Saul}, \bibinfo{person}{Stefan Savage}, {and} \bibinfo{person}{Geoffrey~M.
  Voelker}.} \bibinfo{year}{2009}\natexlab{}.
\newblock \showarticletitle{Beyond Blacklists: Learning to Detect Malicious Web
  Sites from Suspicious URLs}. In \bibinfo{booktitle}{\emph{Proceedings of the
  15th ACM SIGKDD International Conference on Knowledge Discovery and Data
  Mining}}. \bibinfo{pages}{1245--1254}.
\newblock


\bibitem[McGrath and Gupta(2008)]%
        {mcgrath2008}
\bibfield{author}{\bibinfo{person}{D.~K. McGrath} {and} \bibinfo{person}{M.
  Gupta}.} \bibinfo{year}{2008}\natexlab{}.
\newblock \showarticletitle{Behind Phishing: An Examination of Phisher Modi
  Operandi}. In \bibinfo{booktitle}{\emph{Proceedings of the USENIX Workshop on
  Large-Scale Exploits and Emergent Threats (LEET)}}.
\newblock


\bibitem[{National Crime Records Bureau}(2020)]%
        {ncrb}
\bibfield{author}{\bibinfo{person}{{National Crime Records Bureau}}.}
  \bibinfo{year}{2020}\natexlab{}.
\newblock \bibinfo{title}{{Crime in India 2020: Statistics, Volume I}}.
\newblock \bibinfo{howpublished}{\url{https://ncrb.gov.in}}.
\newblock
\newblock
\shownote{Accessed: 2024-01-10}.


\bibitem[O'Harrow(2013)]%
        {o2013zero}
\bibfield{author}{\bibinfo{person}{Robert O'Harrow}.}
  \bibinfo{year}{2013}\natexlab{}.
\newblock \bibinfo{booktitle}{\emph{Zero day: the threat in cyberspace}}.
\newblock \bibinfo{publisher}{Diversion Books}.
\newblock


\bibitem[Patgiri et~al\mbox{.}(2021)]%
        {patgiri2021}
\bibfield{author}{\bibinfo{person}{R. Patgiri} {et~al\mbox{.}}}
  \bibinfo{year}{2021}\natexlab{}.
\newblock \showarticletitle{DeepBF: A Deep Learning based Bloom Filter for
  Malicious URL Detection}.
\newblock \bibinfo{journal}{\emph{IEEE Transactions on Information Forensics
  and Security}}  \bibinfo{volume}{16} (\bibinfo{year}{2021}),
  \bibinfo{pages}{2495--2505}.
\newblock


\bibitem[Rashid et~al\mbox{.}(2024)]%
        {RASHID2024110398}
\bibfield{author}{\bibinfo{person}{Fariza Rashid}, \bibinfo{person}{Ben Doyle},
  \bibinfo{person}{Soyeon~Caren Han}, {and} \bibinfo{person}{Suranga
  Seneviratne}.} \bibinfo{year}{2024}\natexlab{}.
\newblock \showarticletitle{Phishing URL Detection Generalisation Using
  Unsupervised Domain Adaptation}.
\newblock \bibinfo{journal}{\emph{Computer Networks}}  \bibinfo{volume}{245}
  (\bibinfo{year}{2024}), \bibinfo{pages}{110398}.
\newblock
\showISSN{1389-1286}
\href{https://doi.org/10.1016/j.comnet.2024.110398}{doi:\nolinkurl{10.1016/j.comnet.2024.110398}}


\bibitem[Safitra et~al\mbox{.}(2023)]%
        {su151813369}
\bibfield{author}{\bibinfo{person}{Muhammad~Fakhrul Safitra},
  \bibinfo{person}{Muharman Lubis}, {and} \bibinfo{person}{Hanif Fakhrurroja}.}
  \bibinfo{year}{2023}\natexlab{}.
\newblock \showarticletitle{Counterattacking Cyber Threats: A Framework for the
  Future of Cybersecurity}.
\newblock \bibinfo{journal}{\emph{Sustainability}} \bibinfo{volume}{15},
  \bibinfo{number}{18} (\bibinfo{year}{2023}).
\newblock
\showISSN{2071-1050}
\href{https://doi.org/10.3390/su151813369}{doi:\nolinkurl{10.3390/su151813369}}


\bibitem[Sahoo et~al\mbox{.}(2017)]%
        {sahoo2017}
\bibfield{author}{\bibinfo{person}{Doyen Sahoo}, \bibinfo{person}{Chenghao
  Liu}, {and} \bibinfo{person}{Steven~C. Hoi}.}
  \bibinfo{year}{2017}\natexlab{}.
\newblock \showarticletitle{Malicious URL Detection using Machine Learning: A
  Survey}.
\newblock \bibinfo{journal}{\emph{arXiv preprint arXiv:1701.07179}}
  (\bibinfo{year}{2017}).
\newblock
\showeprint[arxiv]{1701.07179}~[cs.CR]


\bibitem[Saleem et~al\mbox{.}(2022)]%
        {saleem2022anonymity}
\bibfield{author}{\bibinfo{person}{Javeriah Saleem}, \bibinfo{person}{Rafiqul
  Islam}, {and} \bibinfo{person}{Muhammad~Ashad Kabir}.}
  \bibinfo{year}{2022}\natexlab{}.
\newblock \showarticletitle{The Anonymity of the Dark Web: A Survey}.
\newblock \bibinfo{journal}{\emph{IEEE Access}}  \bibinfo{volume}{10}
  (\bibinfo{year}{2022}), \bibinfo{pages}{33628--33660}.
\newblock


\bibitem[Schneier(2015)]%
        {schneier2015data}
\bibfield{author}{\bibinfo{person}{Bruce Schneier}.}
  \bibinfo{year}{2015}\natexlab{}.
\newblock \bibinfo{booktitle}{\emph{Data and Goliath: The Hidden Battles to
  Collect Your Data and Control Your World}}.
\newblock \bibinfo{publisher}{W. W. Norton \& Company}.
\newblock


\bibitem[Tajbakhsh et~al\mbox{.}(2020)]%
        {tajbakhsh2020fake}
\bibfield{author}{\bibinfo{person}{Reza Tajbakhsh} {et~al\mbox{.}}}
  \bibinfo{year}{2020}\natexlab{}.
\newblock \showarticletitle{Fake News Detection Based on Hierarchical Attention
  Networks with Semantic--Syntactic Structure}.
\newblock \bibinfo{journal}{\emph{Expert Systems with Applications}}
  \bibinfo{volume}{149} (\bibinfo{year}{2020}), \bibinfo{pages}{113280}.
\newblock


\bibitem[{Verizon}(2025)]%
        {verizon2025dbir}
\bibfield{author}{\bibinfo{person}{{Verizon}}.}
  \bibinfo{year}{2025}\natexlab{}.
\newblock \bibinfo{booktitle}{\emph{2025 Data Breach Investigations Report
  (DBIR)}}.
\newblock \bibinfo{type}{{T}echnical {R}eport}. \bibinfo{institution}{Verizon
  Enterprise Solutions}.
\newblock
\urldef\tempurl%
\url{https://www.verizon.com/business/resources/reports/dbir/}
\showURL{%
\tempurl}
\newblock
\shownote{Accessed: 2025-09-26}.


\bibitem[Yang et~al\mbox{.}(2012)]%
        {yang2012trust}
\bibfield{author}{\bibinfo{person}{Bo Yang}, \bibinfo{person}{Jiming Liu},
  {and} \bibinfo{person}{Irwin King}.} \bibinfo{year}{2012}\natexlab{}.
\newblock \showarticletitle{Trust Inference in Signed Social Networks}. In
  \bibinfo{booktitle}{\emph{Proceedings of the 18th ACM SIGKDD International
  Conference on Knowledge Discovery and Data Mining}}.
  \bibinfo{publisher}{ACM}, \bibinfo{pages}{1048--1056}.
\newblock


\bibitem[Zhou et~al\mbox{.}(2023)]%
        {zhou2023}
\bibfield{author}{\bibinfo{person}{X. Zhou} {et~al\mbox{.}}}
  \bibinfo{year}{2023}\natexlab{}.
\newblock \showarticletitle{Detecting Malicious Domains with Graph Neural
  Networks}.
\newblock \bibinfo{journal}{\emph{IEEE Transactions on Dependable and Secure
  Computing}} (\bibinfo{year}{2023}).
\newblock


\end{thebibliography}

\section*{Author Biographies}

\begin{tabular}{@{}c p{0.85\linewidth}@{}}
\raisebox{-0.95\height}{\includegraphics[width=1in,height=1.25in,clip,keepaspectratio]{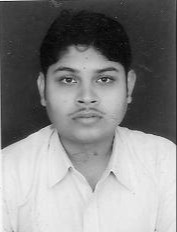}} 
&
\textbf{Avijit Gayen} received the B.Tech. degree in information technology from Kalyani University, Kalyani, India, in 2004, and the M.E. degree in software engineering from Jadavpur University, Kolkata, India, in 2013. From 2013 to 2018, he was a Research Fellow with the Department of Computer Science and Engineering at the Indian Institute of Technology (IIT) Patna, India. He is currently pursuing a Ph.D. degree with the Department of Computer Science and Engineering at the Indian Institute of Information Technology Guwahati (IIITG), India. 
Since 2019, he has been an assistant professor in the Department of Computer Science and Engineering at Techno India University, Kolkata. His research interests include computational social science, network mobility, Federated Learning, Natural Language processing, and related fields. He has contributed to several peer-reviewed publications and projects in these areas.
\end{tabular}
\vspace{.5cm}

\begin{tabular}{@{}c p{0.85\linewidth}@{}}
\raisebox{-0.95\height}{\includegraphics[width=1in,height=1.25in,clip,keepaspectratio]{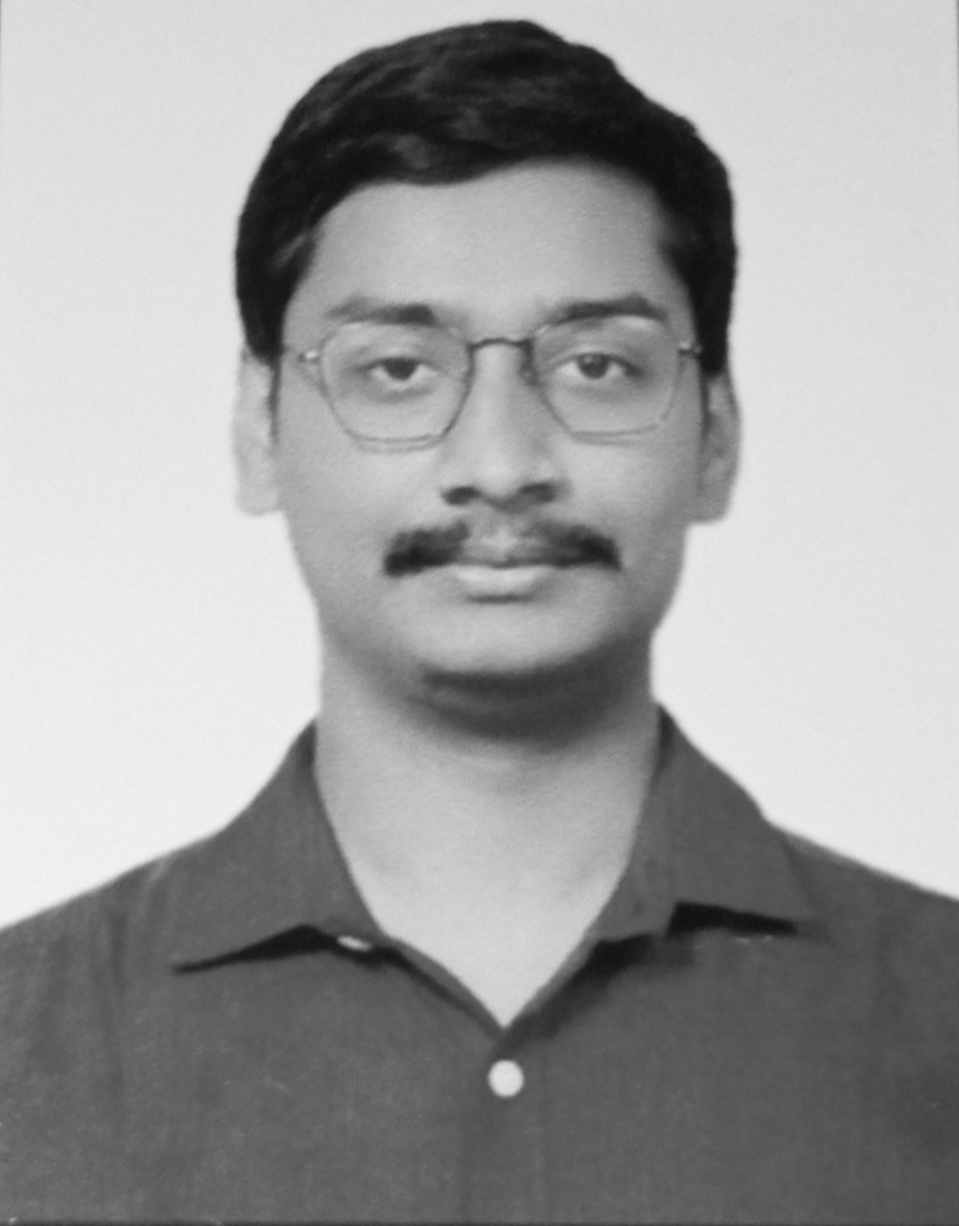}}
&
\textbf{Sayan Mondal}received the B.Tech. degree in Computer Science and Engineering from Techno India University, Kolkata, India, in 2020, and the M.Tech. degree in Computer Software Engineering from Birla Institute of Technology and Science (BITS) Pilani, India, in 2024. He is a Research Scholar (Ph.D.) with the Department of Computer Science and Engineering at the Indian Institute of Information Technology Guwahati (IIITG), India. From November 2020 to March 2025, he was with Virtusa, Hyderabad, India, serving in successive engineering roles. Since March 2025, he has been a Solutions Architect (AI--ML \& ServiceNow) with Humanize, Bengaluru, India. His research interests include cybersecurity, signed networks, graph-based threat detection, trust inference, and applied machine learning systems. He has contributed to open-source projects and industry deployments in phishing detection and LLM-assisted security.
\end{tabular}
\vspace{.5cm}

\begin{tabular}{@{}c p{0.85\linewidth}@{}}
\raisebox{-0.95\height}{\includegraphics[width=1in,height=1.25in,clip,keepaspectratio]{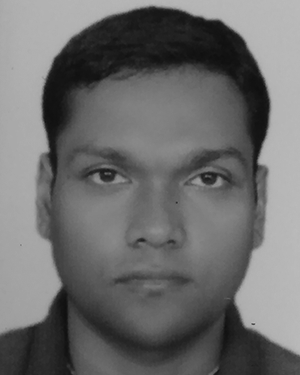}}
&
\textbf{Angshuman Jana} received the B.Tech. degree in Computer Science and Engineering from West Bengal University of Technology, Kolkata, India, in 2011, the M.Tech. degree in Computer Science and Engineering from the National Institute of Technology, Durgapur, India, in 2013, and the Ph.D. degree in Computer Science and Engineering from the Indian Institute of Technology (IIT) Patna, India, in 2019. From January to June 2019, he served as a Visiting Faculty at Motilal Nehru National Institute of Technology (MNNIT), Allahabad. Since July 2019, he has been an assistant professor in the Department of Computer Science and Engineering at the Indian Institute of Information Technology,  Guwahati (IIITG), India. His research interests include formal methods, programming languages, static analysis and verification, information flow security analysis, abstract interpretation, and model checking. He has published several peer-reviewed articles in reputed journals and conferences, particularly focusing on database program verification and secure query analysis.
\end{tabular}










\end{document}